%% file: mismatch_single_letter_upper_bound_paper-ArXiV_March_2021.tex
\documentclass[draftclsnofoot,onecolumn,12pt]{IEEEtran}

\usepackage{amssymb,amsfonts,amsmath,amstext,mathrsfs}
\usepackage{latexsym}
\usepackage{eufrak}
\usepackage{epsfig,psfrag,color,graphics,epsf}
\usepackage{enumerate,cite}
\usepackage{bbm}
\usepackage{stmaryrd}
\usepackage{soul}

\usepackage{amsmath}
\usepackage{amssymb}
\usepackage{amsfonts}
\usepackage{amsthm}
\usepackage{graphicx}
\usepackage{float}
\usepackage{graphics}
\usepackage{graphicx}
\usepackage{euscript}
\usepackage{standalone}
\usepackage{tikz,pgfplots}

\usepackage{psfrag}

\usepackage{enumitem}
\setlist[description]{leftmargin=*}

\DeclareMathOperator*{\argmax}{arg\,max}
\DeclareMathOperator*{\argmin}{arg\,min}

\newtheorem{theorem}{Theorem}
\newtheorem{lemma}{Lemma}

\newtheorem{corollary}{Corollary}
\newtheorem{definition}{Definition}

\newcommand{\dotleq}{%
\DOTSB\mathrel{\mathop{\kern0pt \leq}\limits^{\textstyle.}}}
\newcommand{\dotgeq}{%
\DOTSB\mathrel{\mathop{\kern0pt \geq}\limits^{\textstyle.}}}

\newcommand{\reals} {\mathbb{R}}

\newcommand{\beq} {\begin{equation}}
\newcommand{\eeq} {\end{equation}}
\newcommand{\beqa} {\begin{align}}
\newcommand{\eeqa} {\end{align}}

\newcommand{\indicator}{\mathbbm{1}}

\newcommand {\bx} {\boldsymbol{x}}
\newcommand {\by} {\boldsymbol{y}}
\newcommand {\bz} {\boldsymbol{z}}

\newcommand {\bX} {\boldsymbol{X}}
\newcommand {\bY} {\boldsymbol{Y}}
\newcommand {\bZ} {\boldsymbol{Z}}

\newcommand{\calA}{{\mathcal A}}

\newcommand{\calC}{{\mathcal C}}

\newcommand{\calE}{{\mathcal E}}
\newcommand{\calF}{{\mathcal F}}

\newcommand{\calM}{{\mathcal M}}
\newcommand{\calN}{{\mathcal N}}

\newcommand{\calP}{{\mathcal P}}

\newcommand{\calS}{{\mathcal S}}
\newcommand{\calT}{{\mathcal T}}

\newcommand{\calX}{{\mathcal X}}
\newcommand{\calY}{{\mathcal Y}}
\newcommand{\calZ}{{\mathcal Z}}

\newcommand{\EE}{{\mathbb E}}

\allowdisplaybreaks
\IEEEoverridecommandlockouts

\begin{document}

\title{A Single-Letter Upper Bound on the Mismatch Capacity via Multicast Transmission}

\author{\IEEEauthorblockN{Anelia Somekh-Baruch
\thanks{A.\ Somekh-Baruch is with the Faculty of Engineering at Bar-Ilan University, Ramat-Gan, Israel.  Email: somekha@biu.ac.il. 
  This work was supported by the Israel Science Foundation (ISF), grant no.\ 631/17. This work was accepted (in part) for presentation at the Information Theory Workshop (ITW) 2020. 
}}}

\sloppy

\maketitle
\begin{abstract}
We introduce a new analysis technique to derive a single-letter upper bound on the mismatch capacity of a stationary, single-user, memoryless channel with a decoding metric $q$. Our bound is obtained by considering a multicast transmission over a two-user broadcast channel with decoding metrics $q$ and $\rho$ at the receivers, referred to as $(q,\rho)$-surely degraded. This channel has the property that the intersection event of correct $q$-decoding of receiver $1$ and erroneous $\rho$-decoding of receiver $2$ has zero probability for any fixed-composition codebook of a certain composition $P$. Our bound holds in the strong converse sense of an exponential decay of the probability of correct decoding at rates above the bound. Further, we refine the proof and present a bound that is at least as tight as that of any choice of $\rho$. Several examples that demonstrate the strict improvement of our bound compared to previous results are analyzed. 
Finally, we detect equivalence classes of isomorphic channel-metric pairs $(W,q)$ that share the same mismatch capacity. We prove that if the class contains a matched pair, then our bound is tight and the mismatch capacity of the entire class is fully characterized and is equal to the LM rate, which is achievable by random coding, and may be strictly lower that the matched capacity. 
\end{abstract}

\vspace{6cm}

\pagebreak

\section{Introduction}

One of the most intriguing open problems in Information Theory concerns the fundamental limits of channel coding with a fixed, and possibly suboptimal decoder, where only the codebook can be optimized. 
This problem, termed mismatched decoding, is closely related to other fundamental information-theoretic problems such as the zero-error capacity of the discrete memoryless channel (DMC). 
The question of characterizing the mismatch capacity of a stationary memoryless channel by a single-letter expression (if exists) is a long-standing open problem. 

Achievable rates for channels with mismatched decoding have been studied extensively, especially for DMCs with an additive decoding metric $q$. 
In this case, the decoder's output is the codeword $(x_1,...,x_n)$ which maximizes the accumulated value $\sum_{i=1}^nq(x_i,y_i)$ with the channel output $(y_1,...,y_n)$. 

The simplest lower bound for the mismatch capacity on the DMC $W$ from $\calX$ to $\calY$ and decoding metric $q$ is called the Generalized Mutual Information (GMI), \cite{ShamaiKaplan1993information}. It is achievable by i.i.d.\ random coding, and is given by:
\begin{flalign}\label{eq: CGMI dfn}
R_{q,GMI}(W)=& \max_{Q_X} \underset{\substack{\widetilde{P}_{XY}:\; \widetilde{P}_Y=P_{Y}, \\
\mathbb{E}_{\widetilde{P}}(q(X,Y))\geq \mathbb{E}_P(q(X,Y)) } }{\min}D(\widetilde{P}_{XY}\|Q_X\times P_Y),
\end{flalign}
where $P_{XY}= Q_X\times W$, and $P_Y$ is the marginal $Y$ distribution.

Hui \cite{Hui83} showed that the following rate is achievable using random codes which are nearly constant composition:
\begin{flalign}\label{eq: Cq1 dfn}
R_{q,LM}(W)=& \max_{Q_X} \underset{\substack{\widetilde{P}_{XY}:\; \widetilde{P}_{X}=Q_X,\; \widetilde{P}_Y=P_{Y}, \\
\mathbb{E}_{\widetilde{P}}(q(X,Y))\geq \mathbb{E}_P(q(X,Y)) } }{\min}I_{\widetilde{P}}(X;Y).
\end{flalign}
In \cite{CsiszarNarayan95} it was observed that 
Csisz\'{a}r and K{\"o}rner's \cite{CsiszarKorner81graph} expression for an achievable error-exponent yields $R_{q,LM}(W)$ as an achievable rate.

The rate $R_{q,LM}(W)$ is referred to in the literature as the LM rate, and its multi-letter extension to the channel $W^k$ from $\calX^k$ to $\calY^k$, is also achievable, and in certain cases, it can exceed the LM rate \cite{CsiszarNarayan95}.

Lapidoth \cite{Lapidoth96} introduced an improved lower bound on the mismatch capacity of the DMC by studying the achievable sum-rate of an appropriately chosen mismatched multiple access channel (MAC), whose codebook was obtained by expurgating codewords from the product of the codebooks of the two users. In \cite{SomekhBaruch_mismatchachievableIT2014},\cite{SomekhBaruchISIT_2013} 
a lower bound on the capacity of the single-user channel and an upper bound on its error exponents were derived from the sum-rate of the 
the achievable region and error exponents of a {\it cognitive} MAC using superposition coding or random binning. A refined bound was presented by Scarlett {\it et al.} \cite{ScarlettMartinezGuilleniFabregasISIT_2013} using a refinement of the superposition coding ensemble. For given auxiliary random variables, the results of \cite{SomekhBaruch_mismatchachievableIT2014,SomekhBaruchISIT_2013,ScarlettMartinezGuilleniFabregasISIT_2013} may yield an improvement in the achievable rates of \cite{Lapidoth96} for the DMC.
For other related works and extensions, see the survey on Information-Theoretic foundations of mismatched decoding \cite{ScarlettGuilleniFabregasSomekhBaruchMartinez2020} and references therein, such as 
\cite{Balakirsky_conference_95,ShamaiKaplan1993information,MerhavKaplanLapidothShamai94,Lapidoth96b,GantiLapidothTelatar2000,ShamaiSason2002,ScarlettFabregas2012,ScarlettMartinezGuilleniFabregas_mismatch_2014_IT,
ScalettMartinezGuilleniFabregas_IT_2016,ScarlettPengMerhavMartinezGuilleniFabregas_mismatch_2014_IT}. 

While there have been quite a few works on achievable rates, and some works on multi-letter expressions and upper bounds on the mismatch capacity \cite{SomekhBaruch_general_formula_IT2015} \cite{SomekhBaruchConverses_IT2018}, 
much less is known about single-letter upper bounds. 
Csisz\'{a}r and Narayan \cite{CsiszarNarayan95} proved that a necessary condition for the positivity of the mismatch capacity is the positivity of the LM rate. 
For the binary input binary output case, the mismatch capacity was fully characterized in \cite{CsiszarNarayan95}. They showed that the mismatch capacity $C_q(W)$ is equal to the Shannon capacity, $C(W)$, if $W(0|1)+W(1|0)-1$ and $q(0,1)+q(1,0)-q(0,0)-q(1,1)$ have the same sign, and otherwise $C_q(W)=0$.
The single-letter converse result reported in \cite{Balakirsky95} for binary-input DMCs was disproved in \cite{ScarlettSomehkBaruchMartinezGuilleniFabregas2015}. Specifically, a rate based on superposition coding was shown to exceed the claimed mismatch capacity of \cite{Balakirsky95}.

To the best of our knowledge, the approach of upper bounding the mismatch capacity by an achievable rate of a different channel was first introduced in \cite{SomekhBaruch_general_formula_IT2015}, and later on for the DMC in \cite[Theorem 4]{SomekhBaruchConverses_IT2018}. The latter presented a multi-letter max-min upper bound on $C_q(W)$. 

In a recent work, \cite{Kangarshahi_GuilleniFabregas_ISIT2019,Kangarshahi_GuilleniFabregas_ArXiV_April_2020}, Kangarshahi and Guill\'en i F\`abregas presented a single-letter upper bound on $C_q(W)$, denoted $\bar{R}_q(W)$, for a general DMC $W$ with an additive metric $q$. 
They showed that in certain cases, this bound is strictly lower than the matched capacity, and in the binary-input binary-output case yields the above-mentioned Csisz\'{a}r-Narayan capacity formula. They also proved that if $R >\bar{R}_q(W)$, the maximal error probability converges to $1$ exponentially fast, and presented a numerical algorithm to calculate $\bar{R}_q(W)$. It is also proved that the multi-letter form of the bound is equal to the single-letter bound.

In this paper, we derive a new single-letter upper bound on $C_q(W)$. 
Our bound is based on considering a set of auxiliary broadcast channels that assign zero probability to the intersection event of 
correct $q$-decoding by the $Y$-receiver and erroneous $\rho$-decoding by the $Z$-receiver for every fixed-composition codebook of a certain composition $P$. 
Here, $q$ is the decoding metric of interest, and $\rho$ is some metric 
which can be optimized to yield the tightest bound, including for example the Maximum Likelihood (ML) metric with respect to (w.r.t.)\ the marginal channel to the $Z$-receiver. 

While our bound is always at least as tight as $\bar{R}_q(W)$ (for appropriate choices of $\rho$), 
we show that there are many cases where it is strictly tighter, thereby answering the question that was left open in \cite{Kangarshahi_GuilleniFabregas_ISIT2019,Kangarshahi_GuilleniFabregas_ArXiV_April_2020} as for the tightness of $\bar{R}_q(W)$. 
For example, in the special case of a five-letter noiseless channel ($|\calX|=5$) from $\calX$ to $\calX$, with the pentagon connectivity graph metric which is given by $q_0(x,y)=1$ if $\max_{y} W(y|x)\cdot W(y|x')>0$ and $q_0(x,y)=0$ otherwise, it turns out that  
\begin{equation} \bar{R}_q(W)=C(W)=\log_2(5) [\mbox{bits/channel-use}]\nonumber\end{equation} and we show that our bound $\overline{C}_q(W)$ satisfies
\begin{equation}\overline{C}_q(W)\leq \log_2(5/2) [\mbox{bits/channel-use}].\nonumber\end{equation}
Note that in this case, Lov\'asz \cite{Lovasz1979} established that $C_q(W)=\log_2\sqrt{5}$ [bits/channel-use] (the zero-error capacity of the five-letter typewriter channel).

In addition to providing a tighter bound, our proof is significantly simpler compared to \cite{Kangarshahi_GuilleniFabregas_ISIT2019,Kangarshahi_GuilleniFabregas_ArXiV_April_2020}, and extends verbatim to continuous alphabet stationary memoryless channels (with or without cost constraints), whereas the proof in \cite{Kangarshahi_GuilleniFabregas_ISIT2019,Kangarshahi_GuilleniFabregas_ArXiV_April_2020} relies heavily on the method of types graph theory and holds for the discrete memoryless case only.

Our bounding technique also generalizes Csisz\'{a}r and Narayan's observation that the zero-error capacity $C_0(W)$ of a DMC $W$ is equal to the mismatch capacity of the noiseless channel with input and output alphabets $\calX$ and a decoding metric $q_0$ induced by the connectivity graph associated with that channel. This enables to restate the obvious inequality $C_0(W)\leq C(W)$ as an inequality between two mismatch capacities of two different channels.

Finally, we introduce a relation of superiority between channel-metric pairs, 
and show that it is transitive. We detect an isomorphism between channel-metric pairs, and define equivalence classes of isomorphic pairs. 
 We show that if there exists a matched channel-metric pair $(\widetilde{W},\widetilde{q}_{ML})$, where $\widetilde{q}_{ML}$ is the ML metric w.r.t.\ $\widetilde{W}$ which is isomorphic to $(W,q)$, then 
$C_q(W)=R_{q,LM}(W)=R_{q,GMI}(W)=C(\widetilde{W})$, i.e., the LM rate is equal to the mismatch capacity and it is also equal to the matched capacity of $\widetilde{W}$. 
The existence of an isomorphic matched channel-metric pair is thus a sufficient condition for the tightness of our bound. 
This also yields, as a special case, a sufficient condition for a metric to be capacity-achieving for a certain channel. 
We further extend this notion to isomorphism for a given codebook composition $P$.

This paper is organized as follows. After a short presentation of notational conventions, in Section \ref{sc: Notation}, we present the mismatch decoding problem formally in Section \ref{sc: Statement}.
In Section \ref{sc: first additive}, we present our main results: 
Section \ref{sc: Surely Degraded Broadcast Channels} is devoted to a simple bound, looser than our main result, which holds for additive metrics and stationary memoryless channels. 
Section \ref{sc: first improvement} addresses the multi-letter version of the first bound.
In Section \ref{sc: A Numerical Computation}
we present an algorithm for a numerical computation of the bound.
Section \ref{sc: Channels - Bound Improvement} presents our main bound which is a bound for type-dependent metrics.  
Section \ref{sc: Equivalence Classes of Channel-Metric Pairs} introduces equivalence classes of channel-metric pairs and a sufficient condition for the tightness of our bound.
Section \ref{sc: Examples} presents some examples, and in Section \ref{sc: Conclusion}, we state some concluding remarks. 
Appendix \ref{ap: Spherical Codes}  presents how to adapt our second bound to continuous input alphabet channels with a cost constraint and additive metrics.

 \section{Notation}\label{sc: Notation}

Throughout this paper, scalar random variables are denoted by capital letters, their sample values are denoted by their respective lower case letters, and their alphabets are denoted by their respective calligraphic letters; e.g.\ $X$, $x$, and $\calX$, respectively. A similar convention applies to random vectors of dimension $n$ and their sample values, which are 
denoted in boldface; e.g., $\bx$. The set of all $n$-vectors with components taking values in a certain finite alphabet are denoted by the same alphabet superscripted by n, e.g., $\calX^n$. 
Logarithms are taken to the natural base $e$, unless stated otherwise. 

For a given sequence $\bx \in \calX^n$, where $\calX$ is a finite alphabet,  $\hat{P}_{\bx}$ denotes the empirical distribution on $\calX$ extracted from $\bx$; in other words, $\hat{P}_{\bx}$ is the vector $\{ \hat{P}_{\bx} (x), x\in\calX\}$, where $ \hat{P}_{\bx} (x)$ is the relative frequency of the symbol $x$ in the vector $\bx$. The type-class of $\bx$ is the set of $\bx'\in\calX^n$ such that $\hat{P}_{\bx'}=\hat{P}_{\bx}$, which is denoted $\calT(\hat{P}_{\bx})$. 
The set of all probability distributions on $\calX$ is denoted by $\calP(\calX)$, and the set of empirical distributions of order $n$ on alphabet $\calX$ is denoted $\calP_n(\calX)$.

Information theoretic quantities, such as entropy, conditional entropy, and mutual information are denoted following the usual conventions in the information theory literature, e.g., $H (X )$, $H (X |Y )$, $I(X;Y)$ and so on. To emphasize the dependence of a quantity on a certain underlying probability distribution, say $\mu$, we at times use notations such as $H(\mu )$, $H(\mu_{X|Y})$, $I(\mu_{XY})$, etc. The expectation operator is denoted by $\mathbb{E} (\cdot)$, and  to make the dependence on the underlying distribution $\mu$ clear, it is denoted by $\mathbb{E}_\mu(\cdot)$. The cardinality of a finite set $\calA$ is denoted by $|\calA|$. The indicator function of an event $\calE$ is denoted by $1\{\calE \}$.

For two measures $P,Q$ defined on the same measurable space $(\Omega,\calF)$ the measure $P$ is said to be absolutely continuous w.r.t.\ $Q$ if for every $\calE\in \calF$ such that $Q(\calE)=0$ it also holds that $P(\calE)=0$: this is denoted $P\ll Q$.
The empty set is denoted $\emptyset$. 
 
 \section{Problem Setup}\label{sc: Statement}

Consider transmission over a stationary memoryless channel defined by a conditional probability distribution $W$ from $\calX$ to $\calY$, which are not necessarily finite sets. 
The input-output probabilistic relation is given by:
\begin{flalign}
W^n(\by|\bx) = \prod_{k=1}^n W(y_k|x_k)
\end{flalign}
where $\bx = (x_1,\dotsc,x_n)\in\calX^n$ and $\by= (y_1,\dotsc,y_n)\in\calY^n$ are input and output sequences of length $n$, respectively. Our notation is such that in the discrete case $W(y|x)$ stands for the conditional probability mass function (p.m.f.)\ of $Y$ given $X$, and in the continuous alphabet case, $W(y|x)$ signifies the respective conditional probability density function (p.d.f.)\ 

An encoder maps a message $m\in \{1,\dotsc,M_n\}$ to a channel input sequence $\bx_m\in\calX^n$, creating an $(n,M_n)$-codebook $\calC_n=\{\bx_1,\dotsc,\bx_{M_n}\}$ of rate $R_n=\frac{1}{n}\log M_n$.
The message is a random variable $M$, which is uniformly distributed on $\{1,\dotsc,M_n\}$. 

The decoder's role is to provide an estimate $\hat m \in \{1,\dotsc,M_n\}$ of the transmitted message.
A maximum metric decoder is defined by a function, $q^{}(\bx,\by) : \calX^n\times\calY^n\to \reals$, referred to as ``metric" yielding
\beq\label{eq: 111}
\hat m = \argmax_{i\in \{1,\dotsc,M_n\}} q^{}(\bx_i,\by).
\eeq
If $\hat m\neq m$ an error occurs, and the event of having several maximizers is also considered an error\footnote{Similar to classical channel decoding, breaking ties uniformly at random and declaring an error are equivalent capacity-wise. }. 
The decoder's output, being a function of $\bY$ (the channel output vector of length $n$) is denoted $\hat{M}_q(\bY)$. The resulting average probability of error is given by
\begin{flalign}
P_e(W,\calC_n,q)&= \sum_{m=1}^{M_n}\frac{1}{M_n} W^n(\hat{M}_q(\bY)\neq m|\bX=\bx_m).
\end{flalign}

In this paper, we assume that in the finite-alphabet case, the decoding metric $q^{}(\bx,\by)$ depends on $\bx,\by$ solely via their joint empirical distribution; i.e., $q^{}(\bx,\by)=q(\hat P_{\bx,\by})$, so $q$ can be viewed as a mapping from the empirical distributions to the reals  
$
q:\;\calP_n(\calX\times\calY)\rightarrow\reals$.
More generally, in order not to restrict attention to a specific block-length $n$, we assume that it maps the simplex to a real number; i.e., 
\beq
q:\;\calP(\calX\times\calY)\rightarrow\reals.
\eeq
We refer to this class of metrics as type-dependent (formerly referred to as $\alpha$-decoders by Csisz\'{a}r and K\"{o}rner \cite{CsiszarKorner81graph}).
 In the case of type-dependent metrics, (\ref{eq: 111}) becomes: 
\beq
\hat m = \argmax_{i\in \{1,\dotsc,M_n\}} \,q(\hat{P}_{\bx_i\by}).
\eeq

An important sub-class of type-dependent metrics is the set of additive metrics for which there exists a {\em single-letter} mapping $q:\calX\times\calY\rightarrow \reals$ such that 
\begin{equation}
    q(\hat{P}_{\bx\by}) = \frac{1}{n} \sum_{i=1}^n q(x_i,y_i) = \EE_{\hat{P}_{\bx\by}}[ q(X,Y) ],
    \label{eq:q_additive}
\end{equation}
where for convenience, we slightly abuse notation using $q$ for both the per-letter metric $q(x,y)$ and the $n$-letter metric $q(\hat{P}_{\bx\by})$, since the intention is made clear by the argument of $q(\cdot)$. In this case $q(\hat{P}_{\bx\by})$ is a linear function of $\hat{P}_{\bx\by}$. 
Note that in the special case of equiprobable codewords, the ML decoder, which minimizes the average probability of error, reduces to the additive metric $q(x,y)=\log W(y|x)$. 
For different metrics, the decoder is said to be mismatched \cite{MerhavKaplanLapidothShamai94,CsiszarNarayan95}. 

A rate $R>0$ is said to be achievable for channel $W$ with decoding metric $q$ if for all $\epsilon>0$ there exists a sequence of codes $\{\calC_n\}_{n\in \mathbb{N}}$ such that $|\calC_n|>e^{n(R-\epsilon)}$ and the average probability of error vanishes; i.e., 
$
\lim_{n\rightarrow\infty}P_e(W,\calC_n,q)=0$.

The mismatch capacity of channel $W$ with decoding metric $q$, denoted $C_q(W)$, is the supremum of all achievable rates. 
For brevity, we use the term $q$-mismatch capacity of $W$. The Shannon (matched) capacity of $W$ is denoted $C(W)$. 

We next describe the single-letter bound on $C_q(W)$ of \cite{Kangarshahi_GuilleniFabregas_ISIT2019,Kangarshahi_GuilleniFabregas_ArXiV_April_2020} that was mentioned in the introduction. Let $\calM_{max}(q)$ stand for the following set of joint conditional distributions from $\calX$ to $\calY^2$:
\begin{flalign}\label{eq: calS dfn}
\calM_{max}(q)&=\left\{P_{Y\hat{Y}|X}(y,y'|x) =0 \mbox{ if } x\notin \calS_q(y,y')\right\},
\end{flalign}
where $\calS_q(y,y')=\left\{x:\;x=\argmax_{x'} [q(x',y')-q(x',y)]\right\}$.
The bound on the mismatch capacity $C_q(W)$ of the DMC $W$ with additive metric $q$ is given by: 
\begin{flalign}\label{eq: KG bound}
C_q(W)&\leq \bar{R}_q(W)\triangleq \max_{P_X}\min_{P_{Y'Y|X}\in\calM_{max}(q):\; P_{Y|X}=W} I(X;Y').\end{flalign}
They also showed using the minimax theorem that the min and max are interchangeable, so in fact 
\begin{flalign}\label{eq: KG bound_exp}
\bar{R}_q(W)&= \min_{P_{Y'Y|X}\in\calM_{max}(q):\; P_{Y|X}=W} C(P_{Y'|X}).
\end{flalign}
The proof of this bound relies on constructing a graph in the $\calY^n$ space.

As we shall see, the class of channels that was considered in \cite{Kangarshahi_GuilleniFabregas_ISIT2019,Kangarshahi_GuilleniFabregas_ArXiV_April_2020}   
includes only channels $P_{Y'|X}$ such that $q$-decoding at their output is at least as successful as for the original channel $P_{Y|X}$, for every codebook.

\vspace{1cm}

\section{Main Results}\label{sc: first additive}

In this section, we present a new proof technique that enables us to derive a single-letter upper bound on $C_q(W)$. 
For the ease of presentation, we begin by presenting a simpler bound in Section \ref{sc: Surely Degraded Broadcast Channels}, and in Section \ref{sc: Channels - Bound Improvement} we proceed to our main result. 
Our bounding technique relies on multicast transmission over a broadcast channel $P_{YZ|X}$ from $\calX$ to $\calY\times\calZ$ with the marginal conditional distribution $P_{Y|X}=W$.

\subsection{Surely Degraded Broadcast Channels}\label{sc: Surely Degraded Broadcast Channels}

As stated above, we begin by describing a simple bound which is in fact a corollary of our main result in Theorem \ref{th: ICE and ICE upgraded twice}, but is looser. It holds for additive metrics only.

Let $\calZ$ be a given set (either finite, countably infinite, or continuous), and let $q:\; \calX\times\calY\rightarrow \mathbb{R}$ and $\rho:\; \calX\times\calZ\rightarrow \mathbb{R}$ be two additive metrics. Define 
\begin{flalign}\label{eq: tau dfn }
\tau_{q,\rho}(y,z)=\underset{x'\in\calX}{\max}\;[\rho(x',z)-q(x',y)].
\end{flalign}

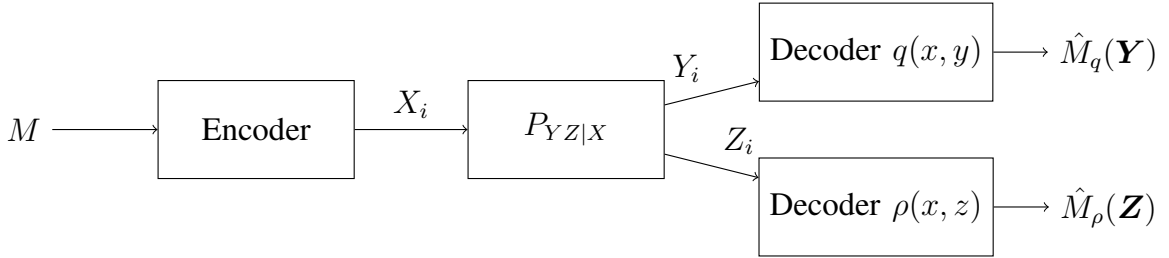
\begin{figure}[H]
	\resizebox{\columnwidth}{!}{\input{Figure_BC_qrho.tex}}
	\caption{A multicast transmission over a broadcast channel with mismatched decoding.}
	\label{BC strongly degraded}
\end{figure}

Consider the following set of broadcast channels\footnote{If the sets $\calX,\calY,\calZ$ are continuous, $\calP(\calY\times\calZ|\calX)$ should be understood as the set of conditional p.d.f.'s rather than conditional p.m.f.'s, and the metrics $\rho$ and $q$ should be such that the resulting support of the distribution $P_{YZ|X}$ is measurable w.r.t.\ the Lesbegue measure.} $P_{YZ|X}\in\calP(\calY\times\calZ|\calX)$ as depicted in Fig.\ \ref{BC strongly degraded}:
\begin{flalign}\label{eq: Gamma q rho dfn}
&\Gamma(q,\rho)
\triangleq \left\{P_{YZ|X}:\;P_{YZ|X}(y,z|x) =0 \;\forall (x,y,z):\; \rho(x,z)-q(x,y)< \tau_{q,\rho}(y,z)\right\}.\end{flalign}
Note that the set $\Gamma(q,\rho)$ may be empty, but at least when $\calZ=\calY$ and $\rho=q$ it contains the channels of the form $P_{YZ|X}=P_{Y|X}\cdot \indicator\{Z=Y\}$.
For reasons that will be clarified later, we refer to channels in $\Gamma(q,\rho)$ as follows. 
\begin{definition}\label{df: sure degradedness} 
We say that the broadcast channel $P_{YZ|X}$ is $(q,\rho)$-surely degraded if $P_{YZ|X}\in \Gamma(q,\rho)$. 
\end{definition}

The first upper bound is given in the following theorem.

 \begin{theorem}\label{th: ICE and ICE upgraded}
For all $\calZ$, additive metrics $q,\rho$,  
and a 
stationary memoryless channel $W$
\begin{flalign}
C_q(W)\leq & \min_{P_{YZ|X}\in\Gamma(q,\rho):\; P_{Y|X}=W}C_{\rho}(P_{Z|X})\label{eq: KG bound2 a}\\
\leq & \min_{P_{YZ|X}\in\Gamma(q,\rho):\; P_{Y|X}=W}C(P_{Z|X})\label{eq: KG bound2}.
\end{flalign}
Further, for all $\epsilon>0$, the average probability of correct decoding of any sequence of codes of rate $R>\min_{P_{YZ|X}\in\Gamma(q,\rho):\; P_{Y|X}=W}C(P_{Z|X})+\epsilon$ vanishes exponentially fast with $n$. 
\end{theorem}
\begin{proof}
Consider a multicast transmission of a single message $M$ over the broadcast channel $P_{YZ|X}\in\Gamma(q,\rho)$ which satisfies $P_{Y|X}=W$. 

By definition of $\Gamma(q,\rho)$, if $P_{YZ|X}(y_i,z_i|x_i)>0$, then 
for all $x'\in\calX$, $\rho(x_i,z_i)-q(x_i,y_i)\geq \rho(x',z_i)-q(x',y_i)$. By additivity of the metrics, and memorylessness of the channel, it follows that 
if $P_{YZ|X}^n(\by,\bz|\bx)>0$, then for every $\bx'\in\calX^n$ it holds that $
\rho(\bx,\bz)-q(\bx,\by)\geq \rho(\bx',\bz)-q(\bx',\by)$. 
Rearranging the inequality we get
\begin{flalign}\label{eq: chain 0}
\rho(\bx,\bz)-\rho(\bx',\bz) \geq q(\bx,\by) -q(\bx',\by),\;\forall (\bx,\by,\bz,\bx'):\; P_{YZ|X}^n(\by,\bz|\bx)>0,
\end{flalign}
where $\bx'\in\calX^n$,
thus, in particular, 
letting $\calC_n=\{\bx_j\}$, $j=1,...,e^{nR}$ be a given codebook, we have that if $P_{YZ|X}^n(\by,\bz|\bx_m)>0$, then
for all $j$,  
\begin{flalign}\label{eq: chain 1}
\rho(\bx_m,\bz)-\rho(\bx_j,\bz) \geq q(\bx_m,\by) -q(\bx_j,\by).
\end{flalign}
Taking the minimum over $j\neq m$ on both sides of the inequality we get that if $P_{YZ|X}^n(\by,\bz|\bx_m)>0$ then
\begin{flalign}\label{eq: chain 2}
\rho(\bx_m,\bz)-\max_{j\neq m} \rho(\bx_j,\bz) \geq q(\bx_m,\by) -\max_{j\neq m}q(\bx_j,\by).
\end{flalign}
This implies that given that $\bx_m$ is transmitted, if the received $\by$ is such that $q(\bx_m,\by)>\max_{j\neq m}q(\bx_j,\by)$ then necessarily also $\bz$ is such that $\rho(\bx_m,\bz)> \max_{j\neq m}q(\bx_j,\bz)$. In words, the error event of the $\rho$-decoder applied to channel output $\bZ$ is contained in the error event of the $q$-decoder applied to channel output $\bY$. This yields
\begin{flalign}\label{eq: SIm2}
\forall n,\;\Pr\left(\hat{M}_q(\bY)= M,\; \hat{M}_{\rho}(\bZ)\neq M\right)=0
\end{flalign}
and consequently
\begin{flalign}
C_{q}(W)\leq \min_{P_{YZ|X}\in\Gamma(q,\rho):\; P_{Y|X}=W}C_{\rho}(P_{Z|X}).
\end{flalign}
Note that for rates exceeding the (potentially) looser upper bound 
$\min_{P_{YZ|X}\in\Gamma(q,\rho):\; P_{Y|X}=W}C(P_{Z|X})$, Eq.\ (\ref{eq: SIm2}) also straightforwardly implies the exponential decay of the probability of correct $q$-decoding at the $\bY$ output, from the strong converse property for the stationary memoryless channel $P_{Z|X}$.
\end{proof}

{\bf \underline{Remarks}}:
Note that any additive $\rho$ gives a valid bound in (\ref{eq: KG bound2 a})-(\ref{eq: KG bound2}), 
so in fact, the bound can be expressed as 
\begin{flalign}
C_q(W)\leq & \min_{(\rho,\; P_{YZ|X}):\; P_{YZ|X}\in\Gamma(q,\rho):\; P_{Y|X}=W}C_{\rho}(P_{Z|X})\label{eq: KG bound2 aabfdi}\\
\leq & \min_{(\rho,\; P_{YZ|X}):\;  P_{YZ|X}\in\Gamma(q,\rho):\; P_{Y|X}=W}C(P_{Z|X})\label{eq: KG bound2sfdbiuvbj}.
\end{flalign}
and in particular, the bound holds for $\rho:\calX\times\calZ\rightarrow \reals$ that may depend on the channel $P_{YZ|X}$; e.g., the matched metric with respect to the marginal $P_{Z|X}$ of $P_{YZ|X}$:
\begin{flalign}
\rho(x,z)=\log P_{Z|X}(z|x)
\end{flalign}
in which case $C_{\rho}(P_{Z|X})$ becomes $C(P_{Z|X})$.

By inspecting (\ref{eq: KG bound}), it is easy to see that the bound $\bar{R}_q(W)$ 
follows from Theorem \ref{th: ICE and ICE upgraded} by taking $\calZ=\calY$ and choosing $\rho=q$, in which case 
$
\calM_{max}(q)= \Gamma(q,q)$,
and by noting that $C_{\rho}(P_{Z|X})\leq C(P_{Z|X})$. 

In Sections \ref{sc: A Numerical Computation} and \ref{sc: Examples}, we provide examples for which the choice $\rho=q$ in (\ref{eq: KG bound2}) is strictly suboptimal; that is, the bound in Theorem \ref{th: ICE and ICE upgraded} is strictly tighter than $\bar{R}_q(W)$ (\ref{eq: KG bound_exp}).

In addition to providing a tighter bound, Theorem \ref{th: ICE and ICE upgraded} has the following advantages over \cite{Kangarshahi_GuilleniFabregas_ISIT2019,Kangarshahi_GuilleniFabregas_ArXiV_April_2020}:
 Our proof is significantly simpler and follows from an observation about multicast transmission over a $(q,\rho)$ surely degraded memoryless broadcast channel. 
 Our proof holds also for continuous alphabet stationary memoryless channels (with or without cost constraints), whereas the proof in \cite{Kangarshahi_GuilleniFabregas_ISIT2019,Kangarshahi_GuilleniFabregas_ArXiV_April_2020} holds for the discrete memoryless case only.

 The term $(q,\rho)$-sure degradedness of Definition \ref{df: sure degradedness} comes from (\ref{eq: SIm2}); i.e., the fact that the error event of the $\rho$-decoder applied to the channel output $\bZ$ is contained in the error event of the $q$-decoder applied to the channel output $\bY$. 

Note that any choice of $\calZ$ and channel $P_{YZ|X}\in\Gamma(q,\rho)$ with marginal $P_{Y|X}=W$ leads to a valid bound, so there are many different (possibly weaker) bounds that are implied by Theorem \ref{th: ICE and ICE upgraded}, without necessarily solving the minimization problem.

If there is any input cost constraint, it should be understood that $C(P_{Z|X})$ and $C_{\rho}(P_{Z|X})$ in (\ref{eq: KG bound2}) are the corresponding capacity and $\rho$-capacity w.r.t\ the cost constraint.

Our bound is clearly tight (and recovers the mismatch capacity formula) for the binary input binary output channel. In other words, in this case the choice $\rho=q$ produces the exact mismatch capacity.

\subsection{A Potentially Improved Bound Using $K$-tuple Additive $\rho$ Metrics}\label{sc: first improvement}

We next address the question whether the multi-letter version of the bound can be tighter than the single-letter bound.

\subsubsection{\ul{A Potential Improvement for $\rho$ that is Additive in $k$-tuples}}

If Theorem \ref{th: ICE and ICE upgraded} is applied to the product channel $W^k$, a valid upper bound on $C_q(W^k)$ is obtained, and therefore also on $C_q(W)$ because $C_q(W)=\frac{1}{k}C_q(W^k)$. 
Let $\rho^{(k)}:\; \calX^k\times \calZ^k\rightarrow \mathbb{R}$ be a $k$-tuple metric which need not necessarily be additive within the $k$-tuples; that is, 
assuming $n$ is an integer multiple of $k$, and denoting $x_i^j=(x_i,...,x_j)$:
\begin{flalign}
\rho(\bx,\bz)&= \frac{1}{n/k}\sum_{i=1}^{n/k} \rho^{(k)}(x_{(i-1)k+1}^{ik} ,z_{(i-1)k+1}^{ik} ).
\end{flalign}  
For simplicity of the notation, for an underlying metric $q:\;\calX\times\calY\rightarrow \mathbb{R}$ we let $q_k$ denote the additive metric $q_k:\;\calX^k\times\calY^k\rightarrow \mathbb{R}$; 
\begin{flalign}
q_k(x_1^k,y_1^k)&= \frac{1}{k}\sum_{i=1}^k q(x_i,y_i).
\end{flalign}  

The upper bound is given in the following corollary of Theorem \ref{th: ICE and ICE upgraded}.

 \begin{corollary}\label{th: Ktuple improvement}
For all $\calZ$, an additive metric $q$, a $k$-tuple metric $\rho^{(k)}$, and a 
stationary memoryless channel $W$
\begin{flalign}
C_q(W)\leq & \min_{P_{Y^kZ^k|X^k}\in\Gamma(q_k,\rho^{(k)}):\; P_{Y^k|X^k}=W^k}\frac{1}{k}C_{\rho^{(k)}}(P_{Z^k|X^k})\label{eq: KG bound2 aalfbvibflb}\\
\leq & \min_{P_{Y^kZ^k|X^k}\in\Gamma(q_k,\rho^{(k)}):\; P_{Y^k|X^k}=W^k}\frac{1}{k}C(P_{Z^k|X^k})\label{eq: KG bound2lfdhviul}.
\end{flalign}
Further, for all $\epsilon>0$, the average probability of correct decoding of any sequence of codes of rate $R>\min_{P_{Y^kZ^k|X^k}\in\Gamma(q_k,\rho^{(k)}):\; P_{Y^k|X^k}=W^k}C(P_{Z^k|X^k})+\epsilon$ vanishes exponentially fast with $n$. 
\end{corollary}
\begin{proof}
The proof is almost identical to the proof of Theorem \ref{th: ICE and ICE upgraded} applied to the product channel $W^k$ with decoding metric $q_k$. The only difference is the fact that $n$ needs to be an integer multiple of $k$. Nevertheless, since this is a converse theorem, the proof holds as is because, by definition of (the mismatch) capacity, it is required that reliable decoding occur for the subsequence of block-lengths $n=k,2k,3k,...$
\end{proof}

The question of whether the bound of Corollary \ref{th: Ktuple improvement} can {\it strictly} improve on the bound of Theorem \ref{th: ICE and ICE upgraded} requires further study.

\subsubsection{\ul{No Improvement for Additive $\rho$}}

We next show that as long as $\rho$ and $q$ are additive metrics, the multi-letter version of the bound (\ref{eq: KG bound2}) cannot improve on the single-letter version.

Let 
\begin{flalign}\label{eq: Gamma q rho dfn k}
&\Gamma^{(k)}(q,\rho)\nonumber\\
&\triangleq \left\{V_{Y^kZ^k|X^k}:\;V(y^k,z^k|x^k) =0 \;\forall x^k,y^k,z^k:\; \sum_{i=1}^k\rho(x_i,z_i)-q(x_i,y_i)< \sum_{i=1}^k\tau_{q,\rho}(y_i,z_i)\right\}.\end{flalign}
\begin{lemma}\label{lm: lemma first gamma}
\begin{flalign}\label{eq: fdkvhfhu}
\frac{1}{k}\min_{P_{Y^kZ^k|X^k}\in\Gamma^{(k)}(q,\rho):\; P_{Y^k|X^k}=W^k}C(P_{Z^k|X^k})
&\geq  \min_{P_{YZ|X}\in\Gamma(q,\rho):\; P_{Y|Z}=W}C(P_{Z|X}) .\end{flalign}
\end{lemma}
The lemma is proved in Appendix \ref{ap: mult no imp}.
In \cite{Kangarshahi_GuilleniFabregas_ISIT2019,Kangarshahi_GuilleniFabregas_ArXiV_April_2020} such claim is proved for $\bar{R}_q(W)$ using the Karush–Kuhn–Tucker (KKT) conditions. 
We present a different proof for our bound, which does not explicitly rely on KKT conditions (and also holds for $\bar{R}_q(W)$ as a special case).

Note Lemma \ref{lm: lemma first gamma} does not rule out the possibility that (\ref{eq: KG bound2lfdhviul}) can improve on the bound (\ref{eq: KG bound2}).
A $\rho$ metric which is additive in pairs, or more generally in $k$-tuples, as suggested in Corollary \ref{th: Ktuple improvement}, could potentially result in an improved bound.
Unlike $\Gamma^{(k)}(q,\rho)$ (see (\ref{eq: Gamma q rho dfn k})), the set $\Gamma(q_k,\rho^{(k)})$ cannot be presented as the $k$-th product of a set of single symbol broadcast channels; that is the inequality
 \begin{flalign}
& \frac{1}{k}\min_{(\rho, P_{Y^kZ^k|X^k}):\; P_{Y^kZ^k|X^k}\in\Gamma^{(k)}(q,\rho):\; P_{Y^k|X^k}=W^k}C(P_{Z^k|X^k})\\
 &\geq 
  \frac{1}{k}\min_{ (\rho^{(k)},P_{Y^kZ^k|X^k}):\;P_{Y^kZ^k|X^k}\in\Gamma(q_k,\rho^{(k)}):\; P_{Y^k|X^k}=W^k}C(P_{Z^k|X^k}),
 \end{flalign}
where $\rho:\; \calX\times\calZ\rightarrow \mathbb{R}$ and $\rho^{(k)}:\; \calX^k\times\calZ^k\rightarrow \mathbb{R}$ may be strict.

\subsection{A Numerical Computation of the Bound of Theorem \ref{th: ICE and ICE upgraded}}\label{sc: A Numerical Computation}

\begin{figure}\label{fig: conversion}
 {\psfrag{Thm1}[][][0.3]{$\bar{R}_q(W)$}
\psfrag{Thm2}[][][0.3]{$\overline{C}_q^{Thm2,\rho}(W)$}
\psfrag{C}[][][0.3]{$C(W)$}
\psfrag{rate}[][][0.5]{$I(t,T)\quad$ [bits/channel-use]}
\centering{\includegraphics[width=\linewidth]{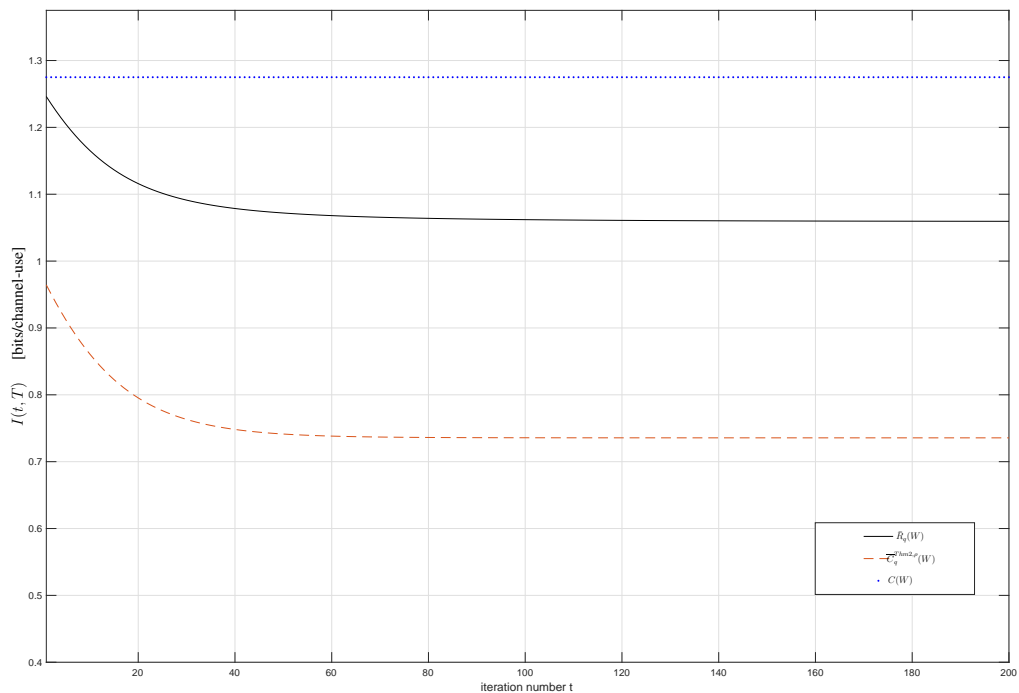}}
\caption{Comparison of the convergence of the numerical calculation of the generalized algorithm of the bound $\overline{C}_q^{Thm2,\rho}(W)$ of Theorem $2$,  and $\bar{R}_q(W)$ with the channel and metrics of (\ref{eq: typetypw wxa11sfbiubdif})-(\ref{eq: typetypw wxa11sfbiubdidkff}) with $T=200$.} }\end{figure}

In the DMC case, the algorithm of \cite{Kangarshahi_GuilleniFabregas_ArXiV_April_2020} for computing $\bar{R}_q(W)$ can be easily modified to compute the bound (\ref{eq: KG bound2}), denoted $\overline{C}_q^{Thm2,\rho}(W)\triangleq \min_{P_{YZ|X}\in\Gamma(q,\rho):\; P_{Y|X}=W}C(P_{Z|X})$.
The generalized algorithm is obtained by substituting $\calS_q(y,y')$ (see definition following Eq.\ (\ref{eq: calS dfn})) by 
\begin{flalign}
\calS_{q,\rho}(y,z)\triangleq \left\{x:\;x=\argmax_{x'} [\rho(x',z)-q(x',y)]\right\}.
\end{flalign} 
Fig.\ \ref{fig: conversion} demonstrates the convergence of the generalized algorithm and displays the approximation at the $t$-th iteration (out of total $T$ iterations), denoted $I(t,T)$, as a function of the iteration number $t$, as well as the approximation of $\bar{R}_q(W)$ for the following channel $W$ and metrics $q,\rho$ where $|\calX|=|\calY|=|\calZ|=5$, 
\begin{flalign}
W(y|x)& =\left\{\begin{array}{ll} 0.76 & x=y\\
0.06 & x\neq y\end{array}\right.\label{eq: typetypw wxa11sfbiubdif}\\
\{q(x,y)\} &=
\left(
\begin{array}{ccccc}
1 & 1 & 0 & 0 & 1\\
1 & 1 & 1 & 0 & 0 \\
0 & 1 & 1 & 1 & 0 \\
0 & 0 & 1 & 1 & 1 \\
1 & 0 & 0 & 1 & 1 
\end{array}\right),\; 
\{\rho(x,z)\} =
\left(
\begin{array}{ccccc}
1 & 1 & 0 & 0 & 0\\
0 & 1 & 1 & 0 & 0 \\
0 & 0 & 1 & 1 & 0 \\
0 & 0 & 0 & 1 & 1 \\
1 & 0 & 0 & 0 & 1 
\end{array}\right).\label{eq: typetypw wxa11sfbiubdidkff}
\end{flalign}
    
\subsection{Surely Degraded Broadcast Channels - Non Rectangular Sets}\label{sc: Channels - Bound Improvement}

In this section, we consider larger sets of broadcast channels compared to $\Gamma(q,\rho)$ that may depend not only on $q$ and $\rho$, but also on the composition $P\in\calP(\calX)$ of the input:
 \begin{flalign}\label{eq: Gamma q rho P dfn}
&\Gamma(q,\rho,P)=\bigg\{P_{YZ|X}:\; 
\max_{V_{XYZ\widetilde{X}}:\;\substack{ V_{XYZ}\ll P\times P_{YZ|X},\\ V_{X}=V_{\widetilde{X}}=P
,\; \rho(V_{XZ})\leq \rho(V_{\widetilde{X}Z})}}\left[q(V_{XY})- q(V_{\widetilde{X}Y})\right]\leq 0\bigg\}\\
&=\bigg\{P_{YZ|X}:\; 
\forall V_{XYZ\widetilde{X}}\mbox{ s.t.\ } \substack{ V_{XYZ}\ll P\times P_{YZ|X},\\ V_{X}=V_{\widetilde{X}}=P
 },\; \rho(V_{XZ})\leq \rho(V_{\widetilde{X}Z})\Rightarrow q(V_{XY})\leq q(V_{\widetilde{X}Y})\bigg\}.
\end{flalign}
Note that in addition to considering a larger set for additive metrics, here we also widen the scope to include $q$ and $\rho$ which are type-dependent metrics, and not necessarily additive, as opposed to Theorem \ref{th: ICE and ICE upgraded}, that holds for additive metrics only. 
An important example for a useful type-dependent metric, which is not additive, is the MMI metric:
\begin{flalign}
q_{{\scriptscriptstyle MMI}}(\hat{P}_{\bx\by})=I(\hat{P}_{\bx\by}),\end{flalign}
 where $I(\hat{P}_{\bx\by})$ is the mutual information induced by the joint distribution $\hat{P}_{\bx\by}$.
 
 Another set of interest is:

 \begin{flalign}\label{eq: Gamma q rho P dfn b rhostaor}
&\Gamma^*(q,P)=\bigg\{P_{YZ|X}:\; 
\max_{V_{XYZ\widetilde{X}}:\;\substack{ V_{XYZ}\ll P\times P_{YZ|X},\\ V_{X}=V_{\widetilde{X}}=P
,\; V_{XZ}=V_{\widetilde{X}Z}}}\left[q(V_{XY})- q(V_{\widetilde{X}Y})\right]\leq 0\bigg\}\\
&=\bigg\{P_{YZ|X}:\; 
\forall V_{XYZ\widetilde{X}}\mbox{ s.t.\ } \substack{ V_{XYZ}\ll P\times P_{YZ|X},\\ V_{X}=V_{\widetilde{X}}=P
 },\; V_{XZ}= V_{\widetilde{X}Z}\Rightarrow q(V_{XY})\leq q(V_{\widetilde{X}Y})\bigg\}.
\end{flalign}
By comparing (\ref{eq: Gamma q rho P dfn}) and (\ref{eq: Gamma q rho P dfn b rhostaor}) it is evident that
\begin{flalign}\label{eq: Gamma q rhoaierj'oijo P dfn b rhostaor}
\Gamma(q,\rho,P)&\subseteq \Gamma^*(q,P),
\end{flalign}
because for every type-dependent metric $V_{XZ}= V_{\widetilde{X}Z}\Rightarrow \rho(V_{\widetilde{X}Z})\Rightarrow q(V_{XY})$.

The following theorem holds in the discrete alphabets case.
\begin{theorem}\label{th: ICE and ICE upgraded twice}
For all finite $\calZ$, type-dependent metrics $q$ and $\rho$,  
and a DMC $W$
\begin{flalign}
C_q(W)&\leq \max_P\min_{P_{YZ|X}:\; P_{YZ|X}\in\Gamma^*(q,P),\; P_{Y|X}=W}I(X;Z)\label{eq: KG bound twiceeugiug}\\
&\leq \max_P\min_{P_{YZ|X}:\; P_{YZ|X}\in\Gamma(q,\rho,P),\; P_{Y|X}=W}I(X;Z).\label{eq: KG bound twice}
\end{flalign}
Further, for all $\epsilon>0$, the average probability of correct decoding of any sequence of codes of rate $R>\max_P\min_{(\rho,P_{YZ|X}):\; P_{YZ|X}\in\Gamma(q,\rho,P):\; P_{Y|X}=W}I(X;Z)+\epsilon$ vanishes exponentially fast with $n$. 
\end{theorem}
The bound (\ref{eq: KG bound twiceeugiug}) has the advantage over (\ref{eq: KG bound twice}) that the set $\Gamma^*(q,P)$ does not depend on a metric $\rho$, and is tighter than for any choice of type-dependent $\rho$. Nevertheless, the bound (\ref{eq: KG bound twice}) can serve to compare between different channel-metrics pairs.

Recall (\ref{eq:q_additive}), it is also easy to see that for any $P$, 
\begin{flalign}
 \Gamma(q,\rho) \subseteq \Gamma(q,\rho,P),
\end{flalign}
 hence, the bound (\ref{eq: KG bound twice}) is tighter than (\ref{eq: KG bound2}) since $C(P_{Z|X})=\max_{P_X} I(X;Z)$, and as we shall see in Section \ref{sc: Examples}, it can be strictly tighter than that of Theorem \ref{th: ICE and ICE upgraded}. This can occur when the maximizing $P$ in (\ref{eq: KG bound twice}) is such that $\Gamma(q,\rho)\subset \Gamma(q,\rho,P)$ and the minimizer $P_{YZ|X}$ in (\ref{eq: KG bound twice}) belongs to $\Gamma(q,\rho,P)\backslash\Gamma(q,\rho)$, in which case the maximization and minimization cannot be commuted. 

As in Section \ref{sc: first improvement}, the multi-letter version of the bound 
\begin{flalign}\label{eq: KG bound twice goung}
C_q(W)\leq &\max_{P_{X^k}\in\calP(\calX^k)}\min_{(\rho^{(k)},P_{Y^kZ^k|X^k}):\; P_{Y^kZ^k|X^k}\in\Gamma(q_k,\rho^{(k)},P_{X^k}),\; P_{Y^k|X^k}=W^k}
\frac{1}{k}I(X^k;Z^k),
\end{flalign}
can potentially yield tighter bounds if we allow $\rho^{(k)}$ to depend on the $k$-th order statistics; that is, if one considers $\rho^{(k)}:\; \calP(\calX^k\times\calZ^k)\rightarrow \mathbb{R}$. 

{\underline{Proof of Theorem \ref{th: ICE and ICE upgraded twice}}:}

We first prove the bound (\ref{eq: KG bound twice}) and then (\ref{eq: KG bound twiceeugiug}). 
Consider transmission of a single message over the stationary memoryless channel $W$. 
Let $\{\calC_n\}$ be a sequence of codebooks at rates $\{R_n\}$, where $R_n>R$ and with vanishing maximal\footnote{Similar to classical channel coding, the mismatch capacity w.r.t.\ average and maximal probabilities of error are equal.} probability of error, $\epsilon_n$. 
Let $P_n$ be a constant composition of a sub-codebook $\widetilde{\calC}_n\subseteq \calC_n$ of rate at least\footnote{This type of sub-codebook always exists because the number of compositions (type-classes) grows polynomially with $n$, and the codebook grows exponentially with $n$.} $R'=R-O(\frac{1}{n}\log n)$. 
Since $\widetilde{\calC}_n\subseteq C_n$, the average probability of error of the sequence of sub-codebooks does not
exceed $\epsilon_n$. 
Now, let $P_{Z|XY}$ be a conditional distribution and let $\rho$ be a type-dependent metric such that the broadcast channel $P_{YZ|X}=W\times P_{Z|XY}$ satisfies $P_{YZ|X}\in\Gamma(q,\rho,P_n)$. 
By definition, since 
$P_{YZ|X}\in\Gamma(q,\rho,P_n)$, 
if $\bx\in\calT(P_n)$ and $P_{YZ|X}^n(\by,\bz|\bx)>0$, then for every $\bx'\in\calT(P_n)$ it holds that $
\rho(\bx',\bz)\geq \rho(\bx,\bz)\Rightarrow q(\bx',\by)\geq q(\bx,\by)$. Similar to (\ref{eq: chain 0})-(\ref{eq: chain 2}) we obtain 
\begin{flalign}\label{eq: chain 2dsiugliug}
\rho(\bx_m,\bz)\leq \max_{j\neq m} \rho(\bx_j,\bz) \Rightarrow  q(\bx_m,\by) \leq \max_{j\neq m}q(\bx_j,\by),
\end{flalign}
and thus for $\bY,\bZ$ the output of the channel $P_{YZ|X}^n$ whose input is uniform over $\widetilde{\calC}_n$, we have
\begin{flalign}\label{eq: SIm2bb}
\forall n,\;\Pr\left(\hat{M}_q(\bY)= M,\; \hat{M}_{\rho}(\bZ)\neq M\right)=0.
\end{flalign}

Let $T$ be a random variable uniformly distributed over $\{1,...,n\}$, independent of $M$, and let $X_T$ be the channel input symbol at time $T$. Since $\widetilde{\calC}_n\subseteq\calT(P_n)$, we have $X_T\sim P_n$. 
Therefore, a standard application of Fano's inequality to channel $P_{Z|X}$ gives
\begin{flalign}\label{eq: Fano hfdfhovad}
R'\leq I(P_n\times P_{Z|X})+\epsilon_n\cdot R'+1/n.
\end{flalign}
Now, since (\ref{eq: Fano hfdfhovad}) holds for all $(\rho,P_{YZ|X}):\; P_{YZ|X}\in\Gamma(q,\rho,P_n)$ such that $P_{Y|X}=W$ this gives
\begin{flalign}
R'\leq  \min_{P_{YZ|X}\in\Gamma(q,\rho,P_n):\; P_{Y|X}=W}I(P_n\times P_{Z|X})+\epsilon_n\cdot R'+1/n
\end{flalign}
and since $P_n\in\calP_n(\calX) \subseteq\calP(\calX)$
\begin{flalign}
C_q(W)\leq \max_{P\in\calP(\calX)} \min_{ P_{YZ|X}\in\Gamma(q,\rho,P):\; P_{Y|X}=W}I(P\times P_{Z|X}).
\end{flalign}

The proof of the claim that for rates exceeding $\max_P\min_{P_{YZ|X}\in\Gamma(q,\rho,P):\; P_{Y|X}=W}I(P\times P_{Z|X})+\epsilon$, the probability of correct $q$-decoding at the $\bY$ output vanishes exponentially fast is deferred to Appendix \ref{ap: ugisugdviugsiug}. 
  
 It remains to prove (\ref{eq: KG bound twiceeugiug}). To this end, we repeat the above proof for the case where the $\bZ$-decoder is genie-aided being informed of the joint empirical distribution of $(\bX,\bZ)$, denoted $\hat{P}_{\bX\bZ}$. In this case, the $\bZ$ decoder errs only if there exists another codeword $\bx'$ of the same joint type-class with $\bZ$ as the transmitted codeword $\bX$. 
Next, by definition, since 
$P_{YZ|X}\in\Gamma^*(q,P_n)$, 
if $\bx\in\calT(P_n)$ and $P_{YZ|X}^n(\by,\bz|\bx)>0$, then for every $\bx'\in\calT(P_n)$ it holds that $\hat{P}_{\bx'\bz}= \hat{P}_{\bx\bz}\Rightarrow q(\bx',\by)\geq q(\bx,\by)$.
Denoting by $\hat{M}(\bZ)$ the output of the genie-aided decoder we obtain
\begin{flalign}\label{eq: SIm2bb}
\forall n,\;\Pr\left(\hat{M}_q(\bY)= M,\; \hat{M}(\bZ)\neq M\right)=0.
\end{flalign}
Next, we show that even with the genie's aid, the rate achievable by the $Z$-decoder is essentially upper bounded by $I(P_n\times P_{Z|X})$. 
By a standard application of Fano's inequality to channel $P_{Z|X}$ gives
\begin{flalign}\label{eq: Fano hfdfhovadskudgvug}
R'&\leq \frac{1}{n}I(\bX;\bZ,\hat{P}_{\bX\bZ})+\epsilon_n\cdot R'+1/n.
\end{flalign}
Now, $\frac{1}{n}I(\bX;\bZ,\hat{P}_{\bX\bZ})\leq \frac{1}{n}I(\bX;\bZ|\hat{P}_{\bX\bZ})+O\left(\frac{\log n}{n}\right)$ because the number of type-classes grows polynomially with $n$, and for all $\epsilon>0$ and $n$ sufficiently large
\begin{flalign}\label{eq: Fano hfdudgvug}
&\frac{1}{n}I(\bX;\bZ|\hat{P}_{\bX\bZ})\nonumber\\
&= \frac{1}{n}H(\bZ|\hat{P}_{\bX\bZ})-\frac{1}{n}H(\bZ|\bX,\hat{P}_{\bX\bZ})\\
&\stackrel{(a)}{\leq} \frac{1}{n}\EE\log|\calT(\hat{P}_{\bZ})|-\frac{1}{n}\EE\log|\calT(\hat{P}_{\bZ|\bX}|\bX)|\\
&\stackrel{(b)}{\leq}I(\hat{P}_{\bX\bZ})+\epsilon ,
\end{flalign}
where $(a)$ follows since conditioning reduces entropy and thus, $H(\bZ|\hat{P}_{\bX\bZ})\leq H(\bZ|\hat{P}_{\bZ})$, and the entropy of a discrete random variable is maximized if it is uniform over its alphabet, in this case $\calT(\hat{P}_{\bZ})$, and thus $
 H(\bZ|\hat{P}_{\bZ})\leq \EE\log|\calT(\hat{P}_{\bZ})|$. Further, since the channel from $\bX$ to $\bZ$ is a DMC, then given $(\bX,\hat{P}_{\bX\bZ})$, $\bZ$ is uniform in $\calT(\hat{P}_{\bZ|\bX}|\bX)$, and thus $H(\bZ|\bX,\hat{P}_{\bX\bZ})=\EE\log|\calT(\hat{P}_{\bZ|\bX}|\bX)|$. 
Step $(b)$ follows since $\frac{1}{n}\log |\calT(\hat{P}_{\bz})|=H(\hat{P}_{\bz})+o(1)$ and similarly $\frac{1}{n}\log |\calT(\hat{P}_{\bz|\bx}|\bx)|=H(\hat{P}_{\bz|\bx})+o(1)$.  Finally, by the law of large numbers, $\hat{P}_{\bX\bZ}$ converges to $P_n\times P_{Z|X}$, and thus by the continuity of the mutual information $\lim_{n\rightarrow\infty }I(\hat{P}_{\bX\bZ})=I(P_n\times P_{Z|X})$. This concludes the proof of (\ref{eq: KG bound twiceeugiug}).

Finally, since
\begin{flalign}
\hat{P}_{\bx_j\bz}= \hat{P}_{\bx_m\bz}\Rightarrow \rho(\hat{P}_{\bx_j\bz})= \rho(\hat{P}_{\bx_m\bz}),
\end{flalign}
it follows from (\ref{eq: chain 2dsiugliug}) that the bound which follows for the genie-aided decoder (\ref{eq: KG bound twiceeugiug}) is tighter than that of any type-dependent $\rho$; i.e., (\ref{eq: KG bound twice}).

 \qed

While the proof of Theorem \ref{th: ICE and ICE upgraded} holds verbatim for continuous input alphabet channels, the proof of Theorem \ref{th: ICE and ICE upgraded twice}, applied to additive metrics, needs to be adapted to the continuous alphabet case. This is because constant composition codebooks are defined for finite alphabets, and the absolute continuity $\ll$ needs to be changed slightly. We describe an application in Appendix \ref{ap: Spherical Codes}.

\subsection{Equivalence Classes of Channel-Metric Pairs}\label{sc: Equivalence Classes of Channel-Metric Pairs}

In this section, we introduce equivalence classes of isomorphic channel-metric pairs $(W,q)$ that share the same mismatch capacity for additive metrics $q$. We prove that if one of the pairs in the class is matched, then the mismatch capacity of the entire class is fully characterized and equal to the LM rate and to the GMI. This gives a sufficient condition for the tightness of our bound. 
In particular, it gives a sufficient condition for a metric to be capacity-achieving. 

Subsequently, we extend this notion to isomorphic channel-metric-composition triplets $(P,W,q)$ where here $q$ can be type-dependent and $P\in\calP_n(\calX)$. 
\begin{definition}\label{df: isomorph}
We say that a channel-metric pair $(P_{Z|X},\rho)$ is superior to the channel-metric pair $(P_{Y|X},q)$ if there exists a joint conditional distribution $P_{YZ|X}\in\Gamma(q,\rho)$, 
whose marginal conditional distributions are $P_{Y|X}$ and $P_{Z|X}$. We say that this channel is surely degraded w.r.t.\ $(q,\rho)$, and we denote the superiority relation by
\begin{flalign}
(P_{Y|X},q)\rightarrowtriangle (P_{Z|X},\rho).
\end{flalign}
If both $(P_{Y|X},q)\rightarrowtriangle (P_{Z|X},\rho)$ and $(P_{Z|X},\rho) \rightarrowtriangle (P_{Y|X},q)$ we say that the pairs are isomorphic and denote this isomorphism relation by 
\begin{flalign}
(P_{Y|X},q)\leftrightarrowtriangle (P_{Z|X},\rho),
\end{flalign}
\end{definition}

The following lemma is proved in Appendix \ref{ap: lfdhblzivufbsiuv}.
\begin{lemma}\label{lm: lfdhblzivufbsiuv}
(a) The relation $\rightarrowtriangle $ is transitive; i.e., if $(W_1,q_1)\rightarrowtriangle (W_2,q_2)$ and $(W_2,q_2)\rightarrowtriangle (W_3,q_3)$ then $(W_1,q_1)\rightarrowtriangle (W_3,q_3)$. 

(b) The relation $\leftrightarrowtriangle $ is an equivalence relation: it is reflexive, symmetric, and transitive.
\end{lemma}

The following theorem provides (among other things) a sufficient condition for the tightness of our bound. In Section \ref{sc: A case of a tight bound}, we present en example, where the bound is provably tight. 
\begin{theorem}\label{lm:ldfslvdfhgivhfiuh}
If $(P_{Y|X},q)\rightarrowtriangle (P_{Z|X},\rho)$ then
\begin{flalign}\label{eq: adhvgiludhviuh}
C_q(P_{Y|X})&\leq C_{\rho}(P_{Z|X}).
\end{flalign}
and consequently, if $(P_{Y|X},q)\leftrightarrowtriangle (P_{Z|X},\rho)$ then
\begin{flalign}\label{eq: trivya}
C_q(P_{Y|X})&=C_{\rho}(P_{Z|X}).
\end{flalign}
If there exists a matched channel-metric pair $(\widetilde{P}_{Z|X},\widetilde{q}_{ML})$ where $\widetilde{q}_{ML}=\log \widetilde{P}_{Z|X}$ is the maximum likelihood metric w.r.t.\ $\widetilde{P}_{Z|X}$ such that $(P_{Y|X},q)\leftrightarrowtriangle (\widetilde{P}_{Z|X},\widetilde{q}_{ML})$ then  
\begin{flalign}
C_q(P_{Y|X})&=R_{q,GMI}(P_{Y|X})=R_{q,LM}(P_{Y|X})=C(\widetilde{P}_{Z|X}),
\end{flalign}
where $R_{q,LM}(P_{Y|X})$ and $R_{q,GMI}(P_{Y|X})$ are the LM and GMI rates of channel $P_{Y|X}$ with decoding metric $q$ (see (\ref{eq: Cq1 dfn}) and (\ref{eq: CGMI dfn})).
\end{theorem}
\begin{proof}
The statement (\ref{eq: adhvgiludhviuh}) follows trivially from Theorem \ref{th: ICE and ICE upgraded}
see (\ref{eq: KG bound2}), and (\ref{eq: trivya}) follows from (\ref{eq: adhvgiludhviuh}). 
The equality $C_q(P_{Y|X})=C(\widetilde{P}_{Z|X})$ is a special case of (\ref{eq: trivya}). It remains to prove $C_q(P_{Y|X})=R_{q,LM}(P_{Y|X})=R_{q,GMI}(P_{Y|X})$. 

Since $(P_{Y|X},q)\leftrightarrowtriangle (\widetilde{P}_{Z|X},\widetilde{q}_{ML})$, similarly to (\ref{eq: chain 0})-(\ref{eq: SIm2}) it follows that there exists a channel $P_{ZY|X}$ with marginals $\widetilde{P}_{Z|X}$ and $P_{Y|X}$ such that one has $\Pr(\hat{M}_{\widetilde{q}_{ML}}(\bZ)=M,\; \hat{M}_q(\bY)\neq M)=0$ for every codebook. Since we also have $C_q(P_{Y|X})=C(\widetilde{P}_{Z|X})$, 
this yields that the random coding scheme, which is capacity-achieving for channel $\widetilde{P}_{Z|X}$, must be capacity-achieving for $P_{Y|X}$ with decoding metric $q$. Since random coding (either i.i.d.\ or constant composition) is capacity-achieving for $\widetilde{P}_{Z|X}$ with $\widetilde{q}_{ML}$, it is also capacity-achieving for $P_{Y|X}$ with $q$-decoding, and thus $C_q(P_{Y|X})=R_{q,LM}(P_{Y|X})=R_{q,GMI}(P_{Y|X})$.
\end{proof}
Note that Theorem \ref{lm:ldfslvdfhgivhfiuh} implies that if $R_{q,GMI}(W)<R_{q,LM}(W)$ then, there exists no matched pair $(\widetilde{P}_{Z|X},\widetilde{q}_{ML})$ such that $(P_{Y|X},q)\leftrightarrowtriangle (\widetilde{P}_{Z|X},\widetilde{q}_{ML})$. 
But, this does not necessarily imply that the bound of Theorem \ref{th: ICE and ICE upgraded} is loose, since there are channel-metric pairs which are not isomorphic but have the same capacity.

The following corollary gives a sufficient condition for a metric $q$ to be capacity-achieving for channel $W$.
\begin{corollary}\label{lm:ldfshgf}
If $(W,\log W)\rightarrowtriangle (W,q)$ then
\begin{flalign}\label{eq: adhvgyrsx}
C_q(W)&= C(W).
\end{flalign}
\end{corollary}

Next, we extend the notion of isomorphism to type-dependent metrics w.r.t.\ the codebook composition $P$. 
\begin{definition}\label{df: isomorph22222}
We say that a composition-channel-metric triplet $(P,P_{Z|X},\rho)$ is superior to the composition-channel-metric triplet $(P,P_{Y|X},q)$ if there exists a joint conditional distribution $P_{YZ|X}\in\Gamma(q,\rho,P)$, 
whose marginal conditional distributions are $P_{Y|X}$ and $P_{Z|X}$. We say that this channel is surely degraded w.r.t.\ $(P,q,\rho)$, and we denote the superiority relation by
\begin{flalign}
(P,P_{Y|X},q)\rightarrowtriangle (P,P_{Z|X},\rho).
\end{flalign}
If both $(P,P_{Y|X},q)\rightarrowtriangle (P,P_{Z|X},\rho)$ and $(P,P_{Z|X},\rho) \rightarrowtriangle (P,P_{Y|X},q)$ we say that the triplets are isomorphic and denote this isomorphism relation by 
\begin{flalign}
(P,P_{Y|X},q)\leftrightarrowtriangle (P,P_{Z|X},\rho),
\end{flalign}
\end{definition}
The following corollary follows by definition of $\Gamma(q,\rho,P)$ from the proof of Theorem \ref{th: ICE and ICE upgraded twice} (see (\ref{eq: chain 2dsiugliug})-(\ref{eq: SIm2bb})).
\begin{corollary}\label{lm:ldfshgfdiuvgusgu}
If there exists a sequence of empirical distributions $P_n\in\calP_n(\calX)$ converging to $P^*$ which is the achiever of $C(W)=\max_{P}I(P \times P_{Y|X})$ such that for all sufficiently large $n$ 
$
(P_n,W,\log W)\rightarrowtriangle (P_n,P_{Z|X},q),
$
then 
\begin{flalign}\label{eq: adhvgyrsxsiufhdv}
C(W)&\leq C_q(P_{Z|X}).
\end{flalign}
In particular, if for all sufficiently large $n$ $(P_n,W,\log W)\rightarrowtriangle (P_n,W,q)$ then 
\begin{flalign}\label{eq: adhvgyrsxsfoihv;oihf}
C_q(W)&= C(W).
\end{flalign}
\end{corollary}
Corollary \ref{lm:ldfshgfdiuvgusgu} gives a sufficient condition (\ref{eq: adhvgyrsxsfoihv;oihf}) which is less strict than that of Corollary \ref{lm:ldfshgf} for a metric to be capacity-achieving.

\subsection{Examples:}\label{sc: Examples}
We next present examples for strict improvements of the bound of Theorem \ref{th: ICE and ICE upgraded} compared to the Shannon capacity, compared to $\bar{R}_q(W)$, and for the improvements of the bound of Theorem \ref{th: ICE and ICE upgraded twice} over that of Theorem \ref{th: ICE and ICE upgraded}, as well as an example of the use of Theorem \ref{lm:ldfslvdfhgivhfiuh} to prove the tightness of our bound.

\subsubsection{\ul{A metric with a degenerate part}}\label{A Metric with a Degenerate Part}

Consider the case where there exist auxiliary alphabets $\calZ,\calA$ and 
a function $\varphi:\; \calZ\times \calA\rightarrow \calY$ such that 
$
q(x,y) =q(x,\varphi(z,a_1))=q(x,\varphi(z,a_2)),\; \forall a_1,a_2\in\calA$.

If $C(P_{Y|X})>C(P_{Z|X})$, then defining $\rho(x,z)\triangleq q(x,\varphi(z,a))$, we obviously obtain
$C_q(P_{Y|X})=C_q(P_{Z,A|X})=C_{\rho}(P_{Z|X})\leq C(P_{Z|X})< C(P_{Y|X})$;
i.e., $C_q(P_{Y|X})\leq C(P_{Z|X})$ is a single-letter bound which is tighter than the Shannon capacity $C(P_{Y|X})$.

\subsubsection{\ul{A noiseless channel with the pentagon graph connectivity metric}}\label{sc: Typewriter Channel}

Consider the noiseless channel $W_r(y|x)=\indicator\{y=x\}$ with $r\triangleq |\calX|$ and the following additive decoding metric
\begin{flalign}
q(x,y)=& \indicator\{(y-x)\mbox{ mod }  r \in\{0,1,r-1\}\}.
\end{flalign}
It is easily verified that the only channel in $\calM_{max}(q)$ that satisfies $P_{Y|X}=W$ is the noiseless channel for which $P_{Y'|XY}(y'|xy)=\indicator\{y'=x\}$.
We demonstrate this for the noiseless channel with the pentagon channel adjacency graph metric; i.e, for $r=5$ with alphabets $\calX=\calY=\{0,1,2,3,4\}$ and with the decoding metric matrix $q(x,y)$ in (\ref{eq: typetypw wxa11sfbiubdidkff}). 
Now, recall the definition of $\calS_q(y,y')$ which appears after (\ref{eq: calS dfn}). 
We have:
\begin{flalign}
\calS_q(y,y')=&\left\{\begin{array}{cl}
\{0,1,2,3,4\} & y=y'\\
 \{2\}& (y,y')=(0,1), (y,y')=(4,3)\\
 \{2,3\} & (y,y')=(0,2), (y,y')=(0,3)\\
  \{3\} & (y,y')=(0,4), (y,y')=(1,2)\\
 \{4\}  & (y,y')=(1,0), (y,y')=(2,3)\\
 \{3,4\}& (y,y')=(1,3), (y,y')=(1,4)\\
 \{0,4\} & (y,y')=(2,0), (y,y')=(2,4)\\
 \{0\}  & (y,y')=(2,1), (y,y')=(3,4)\\
\{0,1\} & (y,y')=(3,0), (y,y')=(3,1)\\
 \{1\} & (y,y')=(3,2),  (y,y')=(4,0)\\
 \{1,2\} & (y,y')=(4,1), (y,y')=(4,2)
 \end{array}
\right.,\end{flalign}
and since the channel dictates $x=y$, we get that $P_{Y'|X}(y'|x)>0$ only for $y'=x$ and thus, $\bar{R}_q(W)$ of (\ref{eq: KG bound}) gives
\begin{flalign}
C_{q}(W)\leq \bar{R}_q(W)=\log_2 5 \,[\mbox{bits/channel-use}]. 
\end{flalign}
It is easy to verify that the typewriter channel
\begin{flalign}\label{eq: ayfdgv}
&
P_{Z|XY}=P_{Z|X}=W_{C_5}\triangleq \frac{1}{2}\cdot \indicator\{(z-x)\mbox{ mod } 5\in\{0,1\}\}
\end{flalign}
with the metric $\rho(x,y)$ in (\ref{eq: typetypw wxa11sfbiubdidkff}).
satisfies the condition that if $P_{YZ|X}(y,z|x)>0$, then $\forall x',\,q(x,y)-q(x',y)\leq  \rho(x,z)-  \rho
(x',z)$; that is
\begin{flalign}\label{eq: example1 efsgv}
(W_5,q)\rightarrowtriangle (W_{C_5},\rho),
\end{flalign}
where, as mentioned previously, $W_5$ is the noiseless channel from $\{0,1,2,3,4\}$ to itself. 
 This is because whenever $P_{YZ|X}(y,z|x)>0$, we have $q(x,y)=\rho(x,y)$ and also $(y,z)=(y,y)$ or $(y,z)=(y,y+1\mod 5)$.
Hence, the condition becomes $\forall x',\,q(x',y)\geq \rho(x',y) $ and $q(x',y)\geq \rho(x',y+1\mod 5) $
which is always satisfied, and therefore
\begin{flalign}
C_{q}(W_5)\leq& C_{\rho}(W_{C_5})=\log_2(5)-\log_2 2\, [\mbox{bits/channel-use}].
\end{flalign}

\normalsize

\subsubsection{\ul{An example with an asymmetric metric}}\label{sc: MATLAB}

Consider the noiseless channel $W_5$ and $q(x,y)=q_{xy}$, $\rho(x,y)=\rho_{xy}$ where
\begin{flalign}\label{eq: dvbkjuhdkvhuyfkuyfu}
\{q_{ij}\}=\left(
\begin{array}{ccccc}
1 & 1 & 0 & 1 & 1\\
1 & 1 & 1 & 0 & 1\\
0 & 1 & 1 & 1 & 1\\
1 & 0 & 1 & 1 & 1\\
1 & 1 & 0 & 1 & 1
\end{array}\right),
\end{flalign}
and $\{\rho_{ij}\}$ as in (\ref{eq: typetypw wxa11sfbiubdidkff}). 
it is easy to verify (e.g.\ with a computer program) that any channel $P_{Z|X}$ whose support is given in the following matrix
\begin{flalign}\label{eq: dvbkjsduhdkvhycjfxjtrx}
\{L_{ij}\}=\left(
\begin{array}{ccccc}
1 & 0 & 0 & 0 & 1\\
1 & 1 & 0 & 0 & 0\\
0 & 0 & 1 & 0 & 0\\
0 & 0 & 0 & 1 & 1\\
1 & 0 & 0 & 1 & 1
\end{array}\right),
\end{flalign}
belongs to $\Gamma(q,\rho)$, and thus satisfies 
\begin{flalign}
(W_5,q)\rightarrowtriangle (W_{Z|X},\rho),\; C_q(W_5)\leq C_\rho(P_{Z|X}).
\end{flalign}

\subsubsection{\ul{A metric with input score}}\label{sc: AX}

Consider additive metrics of the form:
\begin{flalign}
q(x,y)&= a(x,y)+ b(x)\\
\rho(x,y)&= a(x,y)
\end{flalign}
The metrics $\rho$ and $q$ are obviously equivalent for constant composition codebooks since their codewords share the same value of $\sum_{i=1}^n b(x_i)$ value. In fact, this is reflected in Theorem \ref{th: ICE and ICE upgraded twice} which gives $W_{Y|X}\indicator\{Z=Y\}\in \Gamma(q,\rho,P)$, for all $P$, and yields $C_q(W)\leq C_\rho(W)$ (and vice versa, by changing the roles of $\rho$ and $q$).
On the other hand, one can easily find examples for which $\Gamma(a(x,y)+ b(x),a(x,y))=\emptyset$, so Theorem \ref{th: ICE and ICE upgraded} gives meaningless bounds in these cases. For example, taking $a(x,y)\in[0,1]$ for all $x,y$ and $b(x)=0$ for all $x$ except a single symbol $x^*$ for which $b(x^*)=-10$. In this case, $x^*=\argmax_{x'} \rho(x',z)-q(x',y)$, and for $x\neq x^*$, the requirement $P_{YZ|X}(y,z|x)=0$ should hold for all $(y,z)$, which is a requirement that no channel satisfies.

Note that in \cite{Kangarshahi_GuilleniFabregas_ArXiV_April_2020} it was shown that the bound $\bar{R}_q(W)$, is insensitive to metric differences of the form $b(x)$. This is true for $\calM_{max}(q)=\Gamma(q,q)$, but not for $\Gamma(q,\rho)$ in general, and it affects the bound of Theorem \ref{th: ICE and ICE upgraded twice}. Therefore, Theorem \ref{th: ICE and ICE upgraded twice} provides strict improvement over $\bar{R}_q(W)$ and of Theorem \ref{th: ICE and ICE upgraded} in certain cases.

\subsubsection{\ul{Same channel, different metrics}}

(\hspace{1sp}\cite[Example 2]{Lapidoth96}) 
Consider a parallel BSC with $p'<p''<1/2$, and an input alphabet $\{0,1,2,3\}$. The channel transition matrix is given by \begin{flalign}
W_{Y|X}=\left(\begin{array}{cccc}
 (1-p')(1-p'') & (1-p')p'' & p'(1-p'')& p'p''\\
 (1-p')p'' & (1-p')(1-p'') & p'p''&p'(1-p'')\\
p'(1-p'') & p'p''& (1-p')(1-p'') &(1-p')p'' \\
p'p''&p'(1-p'') & (1-p')p'' & (1-p')(1-p'')
\end{array}\right)
\end{flalign}
and $q(x,y)=q_{ML}(x,y)=\log_2 W_{Y|X}(y|x)$, and $\rho(x,y)=r_{x+1,y+1}$ where
\begin{flalign}
  r=\left(\begin{array}{cccc} 0 & -1 & -1 & -2\\ -1 & 0 & -2 & -1\\ -1 & -2 & 0 & -1\\ -2 & -1 & -1 & 0 \end{array}\right).
\end{flalign}
Lapidoth showed that $C_{q_{ML}}(W)= C_\rho(W)$ and the strict inequality $ C_\rho(W)> R_{LM,\rho}(W)$. 

Clearly $C_{q_{ML}}(W)$ is obviously known to equal $C(W)$ because the metric is matched. Nevertheless, it would be interesting to see whether we can show that $C_{q_{ML}}(W)\leq C_\rho(W)$ by treating $q_{ML}$ and $\rho$ in the roles of $q$ and $\rho$ of Theorem \ref{th: ICE and ICE upgraded}.

An exhaustive search of the possibilities (using a simple computer program) reveals that 
one has 

\begin{flalign}
\calS_{q,\rho}(y,z)=\left\{\begin{array}{ll}
\{0\} & (y,z)\in\{(2,0),(3,0),(2,2),(3,2)\}\\
\{1\} & (y,z)\in\{(2,1),(3,1),(2,3),(3,3)\}\\
\{2\} & (y,z)\in\{(0,0),(1,0),(0,2),(1,2)\}\\
\{3\} & (y,z)\in\{(0,1),(1,1),(0,3),(1,3)\}
\end{array}\right.,
\end{flalign}
so evidently, $\Gamma(q,\rho)$ is empty because there is no channel that satisfies this (for example, there is no $z$ such that $P(z|xy)>0$ for $(x,y)=(0,0)$). 
Hence, Theorem \ref{th: ICE and ICE upgraded} does not imply that $C_{q_{ML}}(W)\leq C_\rho(W)$ although there is an equality. Nevertheless, this is not surprising and is consistent with the fact that $ C_\rho(W)> R_{LM,\rho}(W)$; had Theorem \ref{th: ICE and ICE upgraded} been tight in this case, this would have been in contradiction to Theorem \ref{lm:ldfslvdfhgivhfiuh}, which would imply an equality between $C_\rho(W)$ and the LM rate $R_{LM,\rho}(W)$.

\subsubsection{\ul{A case of a tight bound}}\label{sc: A case of a tight bound}

In \cite{Kangarshahi_GuilleniFabregas_ArXiV_April_2020}, the upper bound of (\ref{eq: KG bound}) was computed for the following example
\begin{flalign}\label{eq: typetypw wxa11}
\{W(y|x)\}& =\left(
\begin{array}{ccc}
0.97 & 0.03 & 0 \\
0.1 & 0.1 & 0.8 
\end{array}\right),\;\;\;
\{q(x,y)\} =
\left(
\begin{array}{ccc}
1 & 1 & 1 \\
1 & 1 & 1.36
\end{array}\right),
\end{flalign}
and was shown to equal $0.6182$ [bits/channel-use], where the minimizing channel corresponding to the bound is
\begin{flalign}
P_{Z|X}(z|x)&= \left(
\begin{array}{ccc}
0.5 & 0.5 & 0 \\
0.1 & 0.1 & 0.8 
\end{array}\right).
\end{flalign}
A numerical calculation of the LM rate in \cite{Kangarshahi_GuilleniFabregas_ArXiV_April_2020} suggested that it matches the upper bound. 

One can easily verify that the channel 
$P_{Z|X}$ 
satisfies
\begin{flalign}
(W,q)\leftrightarrowtriangle (P_{Z|X},q) ,\; (P_{Z|X},q)\leftrightarrowtriangle(P_{Z|X},\log P_{Z|X}).
\end{flalign}
Therefore, $(W,q)\leftrightarrowtriangle (P_{Z|X},\log P_{Z|X})$ and the conditions of Theorem \ref{lm:ldfslvdfhgivhfiuh} are satisfied and we have \begin{flalign}
C_q(W)=R_{q,GMI}(W)=R_{q,LM}(W)=C(P_{Z|X}),
\end{flalign}
proving the tightness of the bound (\ref{eq: KG bound}) in this case, and eliminating the need to numerically evaluate the LM rate. 

\section{Conclusion}\label{sc: Conclusion}

In this paper, we presented a single-letter upper bound on the mismatch capacity based on multicast transmission over a broadcast channel. 

The introduction of this multicast transmission setup essentially makes it possible to derive upper bounds quite straightforwardly on the $q$-mismatch capacity of the channel to the first receiver (Channel $1$) by the $\rho$-mismatch capacity of the channel to the second receiver (Channel $2$). While the latter upper bounds the former, the matched capacity of Channel $2$ can be strictly lower than the matched capacity of Channel $1$, thereby yielding a tighter bound compared to the trivial matched capacity of Channel $1$. 
This setup can also be viewed as a generalization of the notion of degradedness of broadcast channels to the mismatched case. 
We further analyzed several examples of channels with mismatched decoding, and demonstrated a strict improvement of our bound compared to previous results in the DMC case, and presented a few examples for continuous alphabet channels as well.

In addition to providing tighter bounds, our method of proof via multicast transmission over a broadcast channel places error events in the same probability space induced by the broadcast channel, yielding a considerably simpler bounding technique compared to that in \cite{Kangarshahi_GuilleniFabregas_ISIT2019,Kangarshahi_GuilleniFabregas_ArXiV_April_2020}. 
Another significant advantage of our approach, is that it holds for continuous alphabet channels verbatim, with or without cost constraints.  
Moreover, our bound holds at greater generality as it also encompasses $q$ and $\rho$ which are type-dependent metrics, and not necessarily additive.
And finally, we presented the bound (\ref{eq: KG bound twiceeugiug}) that is tighter than that of the optimal $\rho$.

The introduction of equivalence classes of channel-metric pairs $(W,q)$, which is important on int own right, enabled us to derive a sufficient condition for the tightness of our bound. This condition states that if the equivalence class includes a matched channel-metric pair, then all the members of that class share the same mismatch capacity, the same LM rate, and the same GMI, which are all equal to the Shannon capacity of the matched pair. The significance of this result is in the fact that it can form a rigorous proof of equality between the numerical computation of the LM rate and our upper bound.

\section{Acknowledgment}
The author is grateful to Jonathan Scarlett, and Sergey Tridenski, for their very helpful and insightful comments and suggestions which improved the quality of this paper.

\appendix

\subsection{Surely Degraded Broadcast Channels w.r.t.\ Spherical Codes}\label{ap: Spherical Codes}

In this section, we demonstrate how to obtain a tighter bound (compared to that of Theorem \ref{th: ICE and ICE upgraded}) for the continuous input alphabet case with a cost constraint using an approach similar to Theorem \ref{th: ICE and ICE upgraded twice}. 
For simplicity of presentation, we consider a power constraint setup in which all the codewords are required to have a fixed (identical) energy\footnote{Shannon \cite{ShannonGaussian1959} showed that in the matched decoding case, considering codebooks of signals of constant energy $n\mathbb{P}^2$ does not reduce the achievable rate compared to $n\mathbb{P}$ expected energy constraint.} equal to $\mathbb{P}$; i.e., they all lie in the  $L_2$ sphere of constant norm $ \sqrt{n\mathbb{P}}$ in $\mathbb{R}^n$, i.e., $\calC_n\subseteq \calS(\mathbb{P})$ where
\begin{flalign}
\calS(\mathbb{P})\triangleq \{\bx\in \mathbb{R}^n:\; ||\bx||_2= \sqrt{n\mathbb{P}}\}.
\end{flalign}

Let $C_q(\mathbb{P},P_{Y|X})$ stand for the $q$-mismatch capacity of channel $P_{Y|X}$ using codebooks that satisfy $\calC_n\subseteq \calS(\mathbb{P})$. 

Theorem \ref{th: ICE and ICE upgraded} tells us that 
\begin{flalign}
C_q(\mathbb{P},P_{Y|X})\leq \min_{P_{YZ|X}\in\Gamma(q,\rho):\; P_{Y|X}=W}C_{\rho}(\mathbb{P},P_{Z|X}).
\end{flalign} Our next theorem shows how this bound can be improved in the spirit of Theorem \ref{th: ICE and ICE upgraded twice}. 

For a distribution $P$, we define $supp(P)$ to be the support of $P$, in the sense that either the probability density function (p.d.f.) or the p.m.f.\ corresponding to $P$ is non-zero. 

We consider the set:
\begin{flalign}
&\Lambda(q,\rho,\mathbb{P})\\
&=\bigg\{V_{YZ|X}:\; 
\forall U_{XYZ\widetilde{X}}\mbox{ s.t.\ } \substack{ supp( U_{XYZ})\subseteq supp( U_X\times V_{YZ|X}),\\ \EE(X^2)=\EE(\widetilde{X}^2)=\mathbb{P}
 },\; \rho(U_{XZ})\leq \rho(U_{\widetilde{X}Z})\Rightarrow q(U_{XY})\leq q(U_{\widetilde{X}Y})\bigg\}\label{eq: laiueghvilugeilugviluegf}
.
\end{flalign}

\begin{theorem}\label{th: ICE and ICE upgraded twice continuous}
For all $\calZ$, additive metric $q$,  
and a 
stationary memoryless channel $W$ 
\begin{flalign}
C_q(\mathbb{P},W)\leq &   \min_{(\rho,P_{YZ|X}):\; P_{YZ|X}\in\Lambda(q,\rho,\mathbb{P}):\; P_{Y|X}=W}C_{\rho}(\mathbb{P},P_{Z|X}).
\\
\leq & \min_{(\rho,P_{YZ|X}):\; P_{YZ|X}\in\Lambda(q,\rho,\mathbb{P}):\; P_{Y|X}=W}\max_{P_X:\; \EE(X^2)=\mathbb{P}}I(X;Z),\label{eq: KG bound twice continuous}
\end{flalign}
where the minimization is over $\rho:\calX\times\calZ\rightarrow \mathbb{R}$, and conditional distributions $P_{YZ|X}$. 
Further, for all $\epsilon>0$, the average probability of correct decoding of any sequence of codes of rate $R> \min_{(\rho,P_{YZ|X}):\; P_{YZ|X}\in\Lambda(q,\rho,\mathbb{P}):\; P_{Y|X}=W}\max_{P_X:\; \EE(X^2)=\mathbb{P}}I(X;Z)+\epsilon$ vanishes exponentially fast with $n$. 
\end{theorem}

{\underline{Proof of Theorem \ref{th: ICE and ICE upgraded twice continuous}}:}
Consider transmission of a single message over the stationary memoryless channel $W$. 
Let $\calC_n=\{\bx_i\}_{i=1}^{e^{nR}}$, where $\calC_n\subseteq\calS(\mathbb{P})$ be a given. 

Now, let $\rho:\calX\times\calZ\rightarrow \mathbb{R}$ be an additive metric, and let $P_{Z|XY}$ be a conditional distribution such that the broadcast channel $P_{YZ|X}=W\times P_{Z|XY}$ satisfies $P_{YZ|X}\in\Lambda(q,\rho,\mathbb{P})$.

Let $\bx,\by,\bz,\widetilde{\bx}$ be given. 
Consider the discrete distribution induced by $\bx,\by,\bz,\widetilde{\bx}$:
\begin{flalign}
U_{XYZ\widetilde{X}}(x,y,z,\widetilde{x})=\frac{1}{n}\sum_{i=1}^n \indicator\{(x_i,y_i,z_i,\widetilde{x}_i)=(x,y,z,\widetilde{x})\},
\end{flalign}
where $x_i,y_i,z_i,\widetilde{x}_i$, are the $i$-th entries of the vectors $\bx,\by,\bz,\widetilde{\bx}$, respectively. 
 Note that  
\begin{itemize}
\item $\bx\in\calS(\mathbb{P})$, $\widetilde{\bx}\in\calS(\mathbb{P})$ can be expressed as  $\EE_{U}(X^2)=\EE_{U}(\widetilde{X}^2)=\mathbb{P}$
\item $(\by,\bz)\in supp ( P_{YZ|X}^n(\cdot|\bx))$ can be expressed as $supp( U_{XYZ})\subseteq supp( U_X\times P_{YZ|X})$
\item $\rho(\widetilde{\bx},\bz)\geq \rho(\bx,\bz)$ can be expressed as $\rho(U_{\widetilde{X}Z})\geq \rho(U_{XZ})$
\item $q(\widetilde{\bx},\by)\geq q(\bx,\by)$ can be expressed as 
$q(U_{\widetilde{X}Y})\geq q(U_{XY})$.
\end{itemize}
Thus, the condition appearing in $\Lambda(q,\rho,\mathbb{P})$ (see (\ref{eq: laiueghvilugeilugviluegf})) is merely a single-letter formulation guaranteeing that 
if $\bx\in\calS(\mathbb{P})$ and $(\by,\bz)$ is a possible channel output in the sense of having positive p.d.f., i.e., $(\by,\bz)\in supp ( P_{YZ|X}^n(\cdot|\bx))$, then for every $\widetilde{\bx}\in\calS(\mathbb{P})$ it holds that 
$
\rho(\widetilde{\bx},\bz)\geq \rho(\bx,\bz)\Rightarrow q(\widetilde{\bx},\by)\geq q(\bx,\by)$.

Since when $\bx_m$ is transmitted, the received signals $(\by,\bz)$ must satisfy $(\by,\bz)\in supp ( P_{YZ|X}^n(\cdot|\bx_m))$, we always have
\begin{flalign}\label{eq: chain 2dsiugliugfgyytfyt}
\rho(\bx_m,\bz)\leq \max_{j\neq m} \rho(\bx_j,\bz) \Rightarrow  q(\bx_m,\by) \leq \max_{j\neq m}q(\bx_j,\by),
\end{flalign}
and thus we obtain (for $\bY,\bZ$ the output of the channel $P_{YZ|X}^n$ whose input is uniform over $\calC_n$),
\begin{flalign}\label{eq: SIm2akdvhiuh}
\Pr\left(\hat{M}_q(\bY)= M,\; \hat{M}_{\rho}(\bZ)\neq M\right)=0
\end{flalign}
and consequently
\begin{flalign}
R\leq C_{\rho}(\mathbb{P},P_{Z|X}),
\end{flalign}
and since this is true for all $(\rho,P_{YZ|X}):\; P_{YZ|X}\in\Lambda(q,\rho,\mathbb{P}):\; P_{Y|X}=W$, we get 
\begin{flalign}
C_{q}(W)\leq \min_{(\rho,P_{YZ|X}):\; P_{YZ|X}\in\Lambda(q,\rho,\mathbb{P}):\; P_{Y|X}=W}C_{\rho}(\mathbb{P},P_{Z|X}).\label{eq: I bound for thodfgjhg}
\end{flalign} 
Since $C_{\rho}(\mathbb{P},P_{Z|X})\leq \max_{P_X:\; \EE(X^2)=\mathbb{P}}I(P_X\times P_{Z|X})$, we obtain the possibly looser upper bound of (\ref{eq: KG bound twice continuous}). And, again, the last assertion of Theorem \ref{th: ICE and ICE upgraded twice continuous} follows from the strong converse for the stationary memoryless channel $P_{Z|X}$.

\qed

As an example, 
%
%
%
%
consider an additive white Gaussian noise (AWGN) channel
\begin{equation}
    Y =  X + N,
\end{equation}
where $N \sim \calN(0,\sigma^2)$ for some noise power $\sigma^2 > 0$, and $X$ and $N$ are independent.  We consider a power constraint corresponding to $\bx_i\in\calS(\mathbb{P})$; that is $\|\bx_i\|^2=\mathbb{P}$ for each codeword $\bx_i$. Note that when matched decoding is concerned, this is equivalent capacity-wise to requiring $\|\bx_i\|^2\leq \mathbb{P}$ \cite{ShannonGaussian1959} (and may be the case with mismatched decoding as well).

Clearly, the maximum likelihood decoding rule is the nearest neighbor rule which minimizes $\|\by - \bx\|_2^2$. 
 Consider a decoder that is based on the wrong assumption that there is a scaling factor $\beta$ (i.e., that the channel is $Y=\beta X +N$), and decodes as output accordingly: 
$    \hat{m} = \argmin_{ j=1,\dotsc,M } \| \by - \beta \bx_j \|_2^2 $. These decoders correspond to the additive decoding metrics $-(y-x)^2$ and $-(y-\beta x)^2$, respectively. Since $\|\by\|^2$ does not affect the decision, we obtain for $\beta>0$
\begin{flalign}
q_1(\bx,\by)&=2<\bx,\by> -\|\bx\|_2^2\\
q_2(\bx,\by)&= 2<\bx,\by> -\beta\|\bx\|_2^2.
\end{flalign}
Now, the mismatched and maximum-likelihood decoding rules are equivalent for codebooks in which all the codewords have the same energy $\|\bx\|_2^2$. Since it is assumed that the codebooks belong to the sphere $\calS(\mathbb{P})$ this is indeed the case.
Nevertheless, in Theorem \ref{th: ICE and ICE upgraded} we have $\Gamma(q_1,q_2)=\emptyset$ where no meaningful result follows, whereas the set $\Lambda(q_1,q_2,\mathbb{P})$ of Theorem \ref{th: ICE and ICE upgraded twice continuous} is non-empty, and in particular, $P_{YZ|X}=P_{Y|X}\cdot \indicator\{Z=Y\}\in \Lambda(q_1,q_2,\mathbb{P})$, so Theorem \ref{th: ICE and ICE upgraded twice continuous} applied to $(q_1,q_2)=(q,\rho)$ and vice versa implies that 
$
C_{q_1}(\mathbb{P},W_{Y|X})= C_{q_2}(\mathbb{P}, W_{Y|X})$.

\subsection{Proof of Lemma \ref{lm: lemma first gamma}:}\label{ap: mult no imp}
 \begin{flalign}
\frac{1}{k}C(P_{Z^k|X^k})
&= \frac{1}{k}\max_{P_{X^k}}I_{P_{X^k}\times P_{Z^k|X^k}}(X^k;Z^k)\\
&\geq \frac{1}{k}\max_{P_X} I_{(P_X)^k\times P_{Z^k|X^k}}(X^k;Z^k)\\
&= \frac{1}{k}\max_{P_X} \sum_{\ell=1}^k I_{(P_X)^k\times P_{Z^k|X^k}}(X_\ell ;Z^k,X^{\ell-1})\label{eq: chaing rule}\\
&\geq \max_{P_{X}}\sum_{\ell=1}^k \frac{1}{k} I_{P_X \times P_{Z_\ell|X_\ell}}(X_\ell;Z_\ell)\label{eq: chaing rule2}\\
&\geq \max_{P_{X}}\min_{\ell} I_{P_X \times P_{Z_\ell|X_\ell}}(X_\ell;Z_\ell)
,\label{eq: guayegfivuahiuhi}
\end{flalign}
where (\ref{eq: chaing rule}) follows from the chain rule for mutual information and the fact that under $(P_X)^k$ the $X_i$'s are i.i.d., so we have $I_{(P_X)^k\times P_{Z^k|X^k}}(X_\ell ;Z^k|X^{\ell-1})=I_{(P_X)^k\times P_{Z^k|X^k}}(X_\ell ;Z^k,X^{\ell-1})$, and where in (\ref{eq: chaing rule}), $P_X \times P_{Z_\ell|X_\ell}$ is the marginal distribution of $(X_\ell,Z_\ell)$ resulting from $(P_X)^k\times P_{Z^k|X^k}$. 

Next, we argue that 
\begin{flalign}
&\max_{P_{X}}\min_{\ell} I_{P_X \times P_{Z_\ell|X_\ell}}(X_\ell;Z_\ell)
\geq \max_{P_{X}}\min_{P_{YZ|X}\in\Gamma(q,\rho):\; P_{Y|Z}=W}I(X;Z).\label{eq: guayegfivuahiuhiuyfkuyfuy}
\end{flalign}
This is because by definition of $\Gamma^{(k)}(q,\rho)$, if $P_{Z^k|X^k}$ is the marginal of $P_{Y^kZ^k|X^k}\in \Gamma^{(k)}(q,\rho)$ such that $P_{Y^k|X^k}=W^k$, it must hold that the marginal $P_{Y_\ell,Z_\ell|X_\ell}$ resulting from $(P_X)^k\times P_{Y^kZ^k|X^k}$ satisfies $P_{Y_\ell|X_\ell}=W$ and lies in $\Gamma(q,\rho)$.
To realize this, note that if $P_{Y^kZ^k|X^k}\in \Gamma^{(k)}(q,\rho)$, then $P_{Y^kZ^k|X^k}(y^k,z^k|x^k)$ can be positive only if 
\begin{flalign}\label{eq: sugfigvziudfghiuv}
\rho(x^k,z^k)-q(x^k,y^k)=\max_{\widetilde{x}^k} [\rho(\widetilde{x}^k,z^k)-q(\widetilde{x}^k,y^k)], 
\end{flalign}
which implies, by the additivity of $q$ and $\rho$, that $P_{Y^kZ^k|X^k}(y^k,z^k|x^k)$ can be positive only if for all $\ell$
\begin{flalign}
\rho(x_\ell,z_\ell)-q(x_\ell,y_\ell)=\max_{\widetilde{x}}[ \rho(\widetilde{x},z_\ell)-q(\widetilde{x},y_\ell)].
\end{flalign}

Now, denote $x^{-\ell}=(x_1,....,x_{\ell-1},x_{\ell+1},...,x_k)$, and note that 
\begin{flalign}
 P_{Y_\ell,Z_\ell|X_\ell}(y,z|x)&= \sum_{y^{-\ell},z^{-\ell},x^{-\ell}}P_X^{k-1}(x^{-\ell}) P_{Y^k,Z^k|X^k}(y^k,z^k|x^k).
\end{flalign}
This implies that if $P_{Y_\ell,Z_\ell|X_\ell}(z|x,y)>0$ there must be at least one triplet $(x^k,y^k,z^k)$ having $(x,y,z)$ as its $\ell$-th entry satisfying $P_{Y^kZ^k|X^k}(y^k,z^k|x^k)>0$. Therefore, if that entry satisfies $\rho(x,z)-q(x,y)<\max_{\widetilde{x}} \rho(\widetilde{x},z)-q(\widetilde{x},y)$, then necessarily $\rho(x^k,z^k)-q(x^k,y^k)<\max_{\widetilde{x}^k} [\rho(\widetilde{x}^k,z^k)-q(\widetilde{x}^k,y^k)]$ and consequently (\ref{eq: sugfigvziudfghiuv}) cannot hold and therefore $P_{Y^kZ^k|X^k}$ cannot be a member of $\Gamma^{(k)}(q,\rho)$.

To conclude, we have 
\begin{flalign}\label{eq: sifugviufhdviuhfivh}
\max_{P_{X}}\min_{P_{YZ|X}\in\Gamma(q,\rho):\; P_{Y|Z}=W}I(X;Z)= \min_{P_{YZ|X}\in\Gamma(q,\rho):\; P_{Y|Z}=W}C(P_{Z|X}),
\end{flalign} which is due to the minimax theorem, which holds since $\{P_{YZ|X}\in \Gamma(q,\rho):\;P_{Y|X}=W\}$ is a convex set and since $I(X;Z)$ is concave in $P_X$ for fixed $P_{Z|X}$ and convex in $P_{Z|X}$ for fixed $P_X$. 
Therefore, we obtain (\ref{eq: fdkvhfhu}). 
\qed

\subsection{Proof of the last claim of Theorem \ref{th: ICE and ICE upgraded twice}:}\label{ap: ugisugdviugsiug}

We prove the claim that for all $\epsilon>0$, the average probability of correct decoding of any sequence of codes of rate $R>\max_P\min_{(\rho,P_{YZ|X}):\; P_{YZ|X}\in\Gamma(q,\rho,P):\; P_{Y|X}=W}I(X;Z)+\epsilon$ vanishes exponentially fast with $n$. 

For a constant composition codebook $\calC_n\subseteq\calT(P)$ the probability of correct decoding (even using ML decoder) over the DMC $P_{Z|X}$, denoted $P_{correct}(\calC_n,P_{Z|X})$, can be upper bounded similarly to \cite[Eq. 6]{DueckKorner1979} as follows:
  \begin{flalign}
  P_{correct}(P,P_{Z|X})\leq (n+1)^{|\calX||\calY|}e^{-n\min_V [D(V\|P_{Z|X}|P)+|R-I(P\times V)|_+]},\nonumber
  \end{flalign}
  where the minimum is over conditional empirical distributions $V$ such that $P\times V\in\calP_n(\calX\times\calY)$.
 If in addition $R>I(P\times P_{Z|X})+\epsilon$, then for $\delta>0$
  \begin{flalign}
 &\min_V [D(V\|P_{Z|X}|P)+|R-I(P\times V)|_+]\nonumber\\
  &\geq \min_V [D(V\|P_{Z|X}|P)+|I(P\times P_{Z|X})+\epsilon-I(P\times V)|_+]\nonumber\\
  &=\min\left\{\substack{
  \min_{V:\; D(V\|P_{Z|X}|P)\geq \delta} [D(V\|P_{Z|X}|P)+\left|I(P\times P_{Z|X})+\epsilon-I(P\times V)\right|_+],\nonumber\\
  \min_{V:\; D(V\|P_{Z|X}|P)\leq \delta} [D(V\|P_{Z|X}|P)+\left|I(P\times P_{Z|X})+\epsilon-I(P\times V)\right|_+]}\right\}
  \nonumber\\
  &\geq \min\left\{
  \delta,
  \min_{V:\; D(V\|P_{Z|X}|P)\leq \delta} [D(V\|P_{Z|X}|P)+\left|I(P\times P_{Z|X})+\epsilon-I(P\times V)\right|_+]\right\}
  \nonumber\\
  &\stackrel{(a)}{\geq}\min\left\{
  \delta,
  \min_{V:\; \frac{1}{2}\|P\times V-P\times P_{Z|X}\|^2\leq \delta} \left|I(P\times P_{Z|X})+\epsilon-I(P\times V)\right|_+\right\}
  \nonumber\\
  &\stackrel{(b)}{\geq}\min\left\{
  \delta,
  \left|\epsilon+4\sqrt{2\delta}\log\frac{\sqrt{2\delta}}{|\calX||\calZ|}\right|_+\right\},\nonumber
  \end{flalign}
  where $\|P_1-P_2\|$ is the total variation distance between $P_1$ and $P_2$, $(a)$ follows from Pinsker's inequality, and $(b)$ follows from \cite[Lemma 2.7]{CsiszarKorner81}, which implies that if $\|P\times V-P\times P_{Z|X}\|\leq d\leq \frac{1}{2}$, then  $\left|I(P\times P_{Z|X})-I(P\times V)\right|\leq -4d\cdot\log\frac{d}{|\calX||\calZ|}$.
  
  Letting $\delta_0\in(0,\frac{1}{2}]$ be such that $\left|4\sqrt{2\delta_0}\log\frac{\sqrt{2\delta_0}}{|\calX||\calZ|}\right|\leq \epsilon/2$, we obtain
  \begin{flalign}
  &\min_V [D(V\|P_{Z|X}|P)+|I(P\times P_{Z|X})+\epsilon-I(P\times V)|_+]\geq \min\left\{
  \delta_0,
  \epsilon/2\right\}.\nonumber
  \end{flalign}
Since any broadcast channel $P_{YZ|X}\in\Gamma(q,\rho,P) $ satisfies that if $\calC_n\subseteq\calT(P)$
then $Pr\left(\hat{M}_q(\bY)= M,\; \hat{M}_{\rho}(\mathbf{Z})\neq M\right)=0$, 
the above argument implies that if $\calC_n\subseteq\calT(P)$ and $\frac{1}{n}\log|\calC_n|=R>\min_{P_{YZ|X}\in\Gamma(q,\rho,P) }I(P\times P_{Z|X})+\epsilon$, then 
\begin{flalign}
\Pr\left(\hat{M}_q(\bY)= M\right)\leq \Pr\left(\hat{M}_\rho(\mathbf{Z})= M\right)\leq P_{correct}(\calC_n,P_{Z|X})\leq e^{-n\min\left\{
  \delta_0,
  \epsilon/2\right\}}.\nonumber
\end{flalign}
Thus, there exists no sequence of constant composition codes of rate 

$R>\max_P \min_{P_{YZ|X}\in\Gamma(q,\rho,P) }I(P\times P_{Z|X})+\epsilon$  with $
\lim_{n\rightarrow\infty }-\frac{1}{n} \log\Pr\left(\hat{M}_q(\bY)= M\right)=0$. 
Since every codebook $\calC_n$ of rate $R$ has at least one constant composition sub-codebook $\calC_n'\subseteq \calC_n$ of rate at least $R-O(\log n)$, the desired result follows. 
\qed

\subsection{Proof of Lemma \ref{lm: lfdhblzivufbsiuv}:}\label{ap: lfdhblzivufbsiuv}

The proof of part (b) is trivial by definition of $\leftrightarrowtriangle $. Hence we only prove (a). 
First note that by definition of $\tau_{q,\rho}(y,z)$ (\ref{eq: tau dfn })
\begin{flalign}\label{eq: tau triv}
\tau_{q_1,q_3}(y_1,y_3)&=\underset{x'\in\calX}{\max}\;[q_3(x',y_3)-q_1(x',y_1)]\\
&= \underset{x'\in\calX}{\max}\;[q_3(x',y_3)-q_2(x',y_2)+q_2(x',y_2)-q_1(x',y_1)]\\
&\leq \tau_{q_2,q_3}(y_2,y_3)+\tau_{q_1,q_2}(y_1,y_2).
\end{flalign}
Next, let $P_{Y_1Y_2|X}\in\Gamma(q_1,q_2)$ with marginals $W_1$ and $W_2$ be given, and let 
$P_{Y_2Y_3|X}\in\Gamma(q_2,q_3)$ with marginals $W_2$ and $W_3$ be given. We need to show that there exists $P_{Y_1Y_3|X}\in \Gamma(q_1,q_3)$ with marginals $W_1$ and $W_3$. 

Let $P_{Y_3|XY_2}$ be the conditional distribution induced by $P_{Y_2Y_3|X}$, and 
consider the channel $P_{Y_1Y_3|X}(y_1,y_3|x)=\sum_{y_2}P_{Y_1Y_2|X}(y_1,y_2|x) \cdot P_{Y_3|XY_2}(y_3|x,y_2)$ which obviously has marginals $W_1$ and $W_3$. 

Now, if $q_3(x,y_3)-q_1(x,y_1)<\tau_{q_1,q_3}(y_1,y_3)$ then from (\ref{eq: tau triv}), for all $y_2\in\calY_2$, 
$q_3(x,y_3)-q_2(x,y_2)+q_2(x,y_2)-q_1(x,y_1)<\tau_{q_1,q_3}(y_1,y_3)\leq \tau_{q_2,q_3}(y_2,y_3)+\tau_{q_1,q_2}(y_1,y_2)$ which implies that either $q_2(x,y_2)-q_1(x,y_1)< \tau_{q_1,q_2}(y_1,y_2)$ or $q_3(x,y_3)-q_2(x,y_2)< \tau_{q_2,q_3}(y_2,y_3)$ which yields 
that either $P_{Y_1Y_2|X}(y_1,y_2|x)=0$ or $P_{Y_2Y_3|X}(y_2,y_3|x)=0$ for all $y_2\in\calY_2$, and consequently 
$P_{Y_1Y_3|X}(y_1,y_3|x)=0$. 
\qed



\end{document}

%% file: Figure_BC_qrho.tex
	\tikzstyle{block} = [draw, fill=white!20, rectangle, 
	minimum height=3em, minimum width=6em]
	\tikzstyle{block2} = [draw, fill=white!20, rectangle, 
	minimum height=3em, minimum width=3em]    
	\tikzstyle{sum} = [draw, fill=white!20, circle, node distance=1cm]
	\tikzstyle{input} = [coordinate]
	\tikzstyle{output} = [coordinate]
	\tikzstyle{pinstyle} = [pin edge={to-,thin,black}]
	
	\begin{tikzpicture}[auto, node distance=2cm]
	\node [input, name=input] {};

	\node [block, right of=input] (Encoder) at (-1.5,0) {Encoder} ;
	\node [block, right of=Encoder, node distance=3cm] (channel) at (1.5,0) {$P_{YZ|X}$};
	\draw [->] (Encoder) -- node[name=u] {$X_i$} (channel);
	\node [block, right of=channel, node distance=3cm] (Decoder) at (5.5,1) {Decoder $q(x,y)$};
	\draw [->] (channel) -- node[name=y] {$Y_i$}  (Decoder) ;  
	\node [output, right of=Decoder] (output) {};
	\node (M) (input)  at (-2.5,0) {$M$};
	\node (M2) (output)  at (11.5,1) {$\hat{M}_q(\bY)$};

	\node [block, right of=channel, node distance=3cm] (Decoder2) at (5.5,-1) {Decoder $\rho(x,z)$};
	\draw [->] (channel) -- node[name=z] {$Z_i$}  (Decoder2) ;  
	\node [output, right of=Decoder2] (output2) {};
	\node (M2) (output2)  at (11.5,-1) {$\hat{M}_{\rho}(\bZ)$};

	\draw [draw,->] (input) -- (Encoder);
	
	\draw [->] (Decoder) -- (output);
	\draw [->] (Decoder2) -- (output2);

%
%
%
%

%
	
	\end{tikzpicture}
	
	